\documentclass[reprint,aps,amsmath,amssymb,superscriptaddress,pra]{revtex4-2}
\usepackage{placeins}
\usepackage{graphicx}
\usepackage{makecell}
\usepackage[T1]{fontenc}
\usepackage[utf8]{inputenc}
\usepackage{amsmath, amssymb, amsfonts}
\usepackage{bm}
\usepackage{physics}
\usepackage{hyperref}
\usepackage{xcolor}
\usepackage{float}
\usepackage{subcaption}
\usepackage{titlesec}
\titleformat{\subsubsection}
  {\normalfont\bfseries\fontsize{8}{10}\selectfont\centering}
  {\thesubsubsection}
  {1em}
  {}

\begin{document}

\title{Strong-Field Coulomb Explosion of Ethane, Propane, and Butane in Circularly Polarized Laser Fields}

\author{Yuxuan Hu}
\affiliation{Department of Physics and Astronomy, Vanderbilt University, Nashville, Tennessee 37235, USA}

\author{Roy Lau}
\affiliation{Department of Physics and Astronomy, Vanderbilt University, Nashville, Tennessee 37235, USA}

\author{Cody Covington}
\affiliation{Department of Chemistry, Austin Peay State University, Clarksville, USA}

\author{Kálmán Varga}
\email{kalman.varga@vanderbilt.edu}
\affiliation{Department of Physics and Astronomy, Vanderbilt University, Nashville, Tennessee 37235, USA}

\begin{abstract}

We investigate the Coulomb explosion of ethane (C$_2$H$_6$), propane (C$_3$H$_8$), and \textit{n}-butane (C$_4$H$_{10}$) driven by 
intense circularly polarized laser pulses using real-time time-dependent density functional theory (RT-TDDFT). The ionization dynamics 
are benchmarked against those obtained with linearly polarized fields oriented along the $x$, $y$, and $z$ axes at the same 
peak intensity. Under the laser conditions considered here, circular polarization produces greater ionization than any of the 
linearly polarized configurations for all three molecules, indicating that the rotating electric field enhances the initial electron-removal stage that triggers Coulomb explosion.
Using circularly polarized excitation, we systematically characterize fragmentation thresholds, product distributions, channel branching ratios, 
and bond-breaking dynamics across the alkane series. Atomic hydrogen is the most abundant fragment in all three systems, 
demonstrating that hydrogen loss is the dominant fragmentation pathway. Ethane primarily retains its two-carbon backbone through 
partial dehydrogenation, propane exhibits the broadest range of fragmentation channels and the strongest competition between 
C--H and C--C bond cleavage within the present ensemble, and butane favors backbone cleavage into relatively stable two-carbon fragments, most notably through the $2\mathrm{C_2H_4} + 2\mathrm{H}$ channel. Analysis of the earliest bond-breaking events further shows that 
C--H dissociation is the preferred initial fragmentation step throughout the series, although the degree of competition with C--C cleavage depends on molecular size.
Overall, these results show that circular polarization influences Coulomb explosion at multiple interconnected stages, including ionization efficiency, fragmentation 
threshold, product distribution, and bond-breaking sequence. This work provides a foundation for future investigations of 
polarization-dependent strong-field fragmentation in larger polyatomic molecules.

\end{abstract}

\maketitle

\section{Introduction}

Coulomb explosion is a strong-field process in which an intense
ultrafast laser
pulse rapidly strips multiple electrons from a molecule, leaving
behind a highly
charged ionic framework that disintegrates under strong internal Coulomb
repulsion. The resulting fragment distributions encode information about
molecular structure, charge redistribution, and laser--matter
coupling, making
Coulomb explosion a powerful tool for probing fragmentation pathways and
ionization dynamics in polyatomic systems. It has found broad
applications in
structural-dynamics
studies~\cite{Allum2018,Endo2020,Erattupuzha2017,Legare2006,Stapelfeldt1995,
Hu2020,Zhang2024,Dantus2024,Champenois2023,Crane2023PCCP,Ekanayake2017,
Ekanayake2018}, the generation of bright keV X-ray
photons~\cite{Ditmire1995,McPherson1994}, molecular
imaging~\cite{Vager1989,Schouder2022,Li2022,Herwig2013,Unwin2023,Mogyorosi2020,
Pitzer2013,Slater2015,Boll2022,Singh2024,Zhou2020,Yatsuhashi2018,Wu2011,
Minion2022,Crane2021,Hishikawa2007,Ma2019,Corrales2019,Bittner2022,Zhang2022,
Kranabetter2024,Kristensen2023,Rolles2023,Bhattacharyya2022,Cheng2023,
Howard2023}, and the production of highly energetic
electrons~\cite{Shao1996}. Fragment formation in Coulomb explosion has
been
extensively explored across a broad range of molecular
systems~\cite{Bhattacharyya2022,Crane2023PCCP,Hishikawa1998,Pitzer2016,
Luzon2019,Cornaggia1992,Wu2019,Xu2009,Xu2010,Kraus2011,Severt2024,Kwon2023,
Michie2019}.

In the variant known as Coulomb explosion imaging (CEI), the momenta
of ionic
fragments---measured in coincidence whenever possible---provide direct
access to
molecular geometry, rotational distributions, and fragmentation
pathways at the
instant of ionization. CEI has therefore been widely used as a
``molecular
camera'' for static structural determination, time-resolved
photodissociation
dynamics, and strong-field fragmentation
studies~\cite{Crane2023PCCP,Hering1999}.

Comparing linear and circular polarization is important because the
two field geometries couple to molecular electron dynamics in
fundamentally different ways. In a linearly
polarized
field, the electric field oscillates along a fixed axis, and
strong-field
ionization is commonly interpreted through recollision-driven
dynamics: one
electron is first tunnel-ionized and, after the field reverses, may be
driven
back toward the parent ion, where it can trigger excitation or secondary
ionization~\cite{Corkum1993,Corkum2011}. In a circularly polarized
field, by
contrast, the electric-field vector rotates continuously in the
polarization
plane, guiding emitted electrons along curved trajectories that
generally do not
return to
the parent ion. Circular polarization is therefore generally expected to
suppress recollision and to alter both the efficiency and the
character of
strong-field ionization.

Experimental and theoretical evidence, however, reveals a more
nuanced
picture. Xie~\textit{et~al.}~\cite{Xie2014} observed that high-energy
$\mathrm{H^+}$ fragments in acetylene were strongly suppressed under
circular
polarization, an effect attributed to reduced nonsequential double
ionization.
Yet both experiment and theory have shown that nonsequential double
ionization is
not entirely absent in circularly polarized
fields~\cite{Guo1998,Tong2013}.
These findings indicate that circular polarization does not simply
reduce
ionization yields; rather, it reshapes the interplay among sequential
ionization, nonsequential ionization, charge redistribution, and nuclear
dynamics in ways that depend sensitively on the molecular system.
Consistent
with this view, early CEI measurements on $\mathrm{N_2}$ and
$\mathrm{CO_2}$
showed that kinetic-energy release can be comparable for linear and
circular
polarization under certain conditions, indicating that explosion
dynamics are
not governed solely by the instantaneous field
direction~\cite{Hering1999}.

Circularly polarized pulses have also proven valuable as probe pulses in
time-resolved CEI schemes, where a pump pulse first prepares rotational,
vibrational, or dissociative dynamics and a delayed circularly
polarized probe
maps the evolving molecular structure onto fragment momenta. This
approach has
enabled direct imaging of rotational wave-packet revivals in diatomic
molecules~\cite{Dooley2003} and has been extended to cluster systems,
where CEI
reveals how a weakly bound neighbor modifies molecular
rotation~\cite{Lu2024}. More broadly, circular polarization provides
access to
molecular-frame photoelectron distributions and helicity-dependent
fragmentation
observables~\cite{Holmegaard2010,Spanner2012,Fehre2019,Fujise2022}. A
quantitative interpretation of these measurements requires
calculations of
channel-resolved ionization probabilities, charge-state evolution,
electron
trajectories, and Coulomb-trajectory propagation, so as to determine
whether
observed fragment momenta primarily reflect the initial molecular
geometry,
polarization-dependent ionization dynamics, or nuclear rearrangement
during the
buildup of charge.

Hydrocarbon molecules have served as important model systems in Coulomb
explosion
experiments~\cite{Roither2011,Markevitch2004,Cornaggia1995,Shimizu2000,
Palaniyappan2010}. Recent TDDFT studies have examined Coulomb
explosion in
hydrocarbons under linearly polarized fields, while RT-TDDFT
simulations of
circularly polarized excitation have been largely confined to smaller
systems
such as $\mathrm{H_2}$ and $\mathrm{C_2H_2}$, with emphasis on alignment
effects and enhanced ionization~\cite{Taylor2025,Russakoff2015}. The
response
of the small alkane series---ethane ($\mathrm{C_2H_6}$), propane
($\mathrm{C_3H_8}$), and \textit{n}-butane ($\mathrm{C_4H_{10}}$)---to
circularly polarized fields has not yet been systematically
characterized in
terms of polarization-dependent ionization, fragmentation thresholds,
product
distributions, or bond-breaking order.

To address this gap, the present work uses RT-TDDFT to investigate
the Coulomb
explosion of ethane, propane, and \textit{n}-butane driven by circularly
polarized laser pulses. To isolate the effect of the rotating field
geometry,
we first compare ionization dynamics produced by a circularly
polarized field in
the $xy$ plane with those driven by linearly polarized fields along
the $x$, $y$,
and $z$ directions at the same peak intensity. We then characterize the
fragmentation dynamics under circular polarization by determining
fragmentation
thresholds, product distributions, dominant branching channels, and
the timing
of the first C--H and C--C bond-breaking events. The temporal
correlation between
the first C--H and C--C breaking events is further analyzed to
establish the
preferred microscopic sequence of fragmentation. In this way, the
study addresses
both how circular polarization modifies the early electron-removal
stage that
initiates Coulomb explosion and how it governs the subsequent nuclear
breakup
dynamics in small hydrocarbon molecules.

\section{Computational Details}

The simulations were performed using time-dependent density functional theory
(TDDFT) to model the electron dynamics on a real-space grid with real-time
propagation~\cite{Runge1984}. Core electrons, which are computationally expensive to treat explicitly, were described using norm-conserving Troullier--Martins pseudopotentials~\cite{Troullier1991}. The time-dependent Kohn--Sham (KS) Hamiltonian is written as
\begin{equation}
\begin{aligned}
\hat{H}_{\mathrm{KS}}(t)
&=
-\frac{\hbar^2}{2m}\nabla^2
+
V_{\mathrm{ion}}(\mathbf{r},t)
+
V_{\mathrm{H}}[\rho](\mathbf{r},t) \\
&\quad
+
V_{\mathrm{xc}}[\rho](\mathbf{r},t)
+
V_{\mathrm{laser}}(\mathbf{r},t).
\end{aligned}
\label{eq:HKS}
\end{equation}

Here, \(\rho(\mathbf{r},t)\) is the electron density, defined as the sum over all
occupied orbitals,
\begin{equation}
\rho(\mathbf{r},t)
=
\sum_{k=1}^{N_{\mathrm{orbitals}}}
2\,|\psi_k(\mathbf{r},t)|^2,
\label{eq:density}
\end{equation}
where the factor of \(2\) accounts for spin degeneracy, and \(k\) labels the
occupied Kohn--Sham orbitals.

The ionic potential \(V_{\mathrm{ion}}\) in Eq.~(\ref{eq:HKS}) is represented using
norm-conserving pseudopotentials centered at each ion. The Hartree potential
\(V_{\mathrm{H}}\) accounts for the electrostatic electron--electron interaction and
is given by
\begin{equation}
V_{\mathrm{H}}(\mathbf{r},t)
=
\int
\frac{\rho(\mathbf{r}',t)}{|\mathbf{r}-\mathbf{r}'|}
\,d\mathbf{r}'.
\label{eq:Hartree}
\end{equation}
The term \(V_{\mathrm{xc}}\) denotes the exchange-correlation potential, whose
exact form is a complicated functional of the full time-dependent electron
density. In the present work, it is approximated within the adiabatic
local-density approximation (ALDA) using the parameterization of Perdew and
Zunger~\cite{Perdew1981}. 

The interaction with the laser field is described in the dipole approximation,
\begin{equation}
V_{\mathrm{laser}}(\mathbf{r},t)
=
\mathbf{r}\cdot \mathbf{E}_{\mathrm{laser}}(t).
\label{eq:Vlaser}
\end{equation}

Since the present work uses circularly polarized light, the laser electric field
contains two orthogonal components with a phase difference of \(\pi/2\). It is
written as
\begin{equation}
\mathbf{E}_{\mathrm{laser}}(t)
=
E_{\max}
\exp\!\left[
-\frac{(t-t_0)^2}{2a^2}
\right]
\left[
\sin(\omega t)\,\hat{\mathbf{x}}
+
\sin(\omega t+\phi)\,\hat{\mathbf{y}}
\right],
\label{eq:circularfield}
\end{equation}
where \(E_{\max}\) is the peak field amplitude, \(t_0\) is the center of the pulse,
\(a\) determines the Gaussian envelope width, \(\omega\) is the laser frequency,
and \(\phi\) is the phase difference between the two orthogonal components. For
the present  circularly polarized pulse, we use \(\phi=\pi/2\). The
laser wavelength is \(890~\mathrm{nm}\), and the pulse width is \(9.18~\mathrm{fs}\).
This corresponds to left-handed circularly polarized light. Since the
molecules investigated in this study are achiral, both left- and
right-handed circular polarizations would yield identical results.
The laser wavelength and pulse width are chosen to match the
experimental conditions used in prior linearly polarized Coulomb
explosion studies of these systems \cite{Taylor2025}, ensuring that the circularly
polarized results presented here are directly comparable to the
existing experimental baseline.

At the beginning of the TDDFT calculation, the ground state of the molecule is
obtained from a static density-functional theory calculation. The resulting
Kohn--Sham orbitals are then propagated in real time according to the
time-dependent Kohn--Sham equation
\begin{equation}
i\hbar\frac{\partial \psi_k(\mathbf{r},t)}{\partial t}
=
\hat{H}_{\mathrm{KS}}(t)\psi_k(\mathbf{r},t).
\label{eq:TDKS}
\end{equation}

Equation~(\ref{eq:TDKS}) is solved using the time propagator
\begin{equation}
\psi_k(\mathbf{r},t+\delta t)
=
\exp\!\left(
-\frac{i\hat{H}_{\mathrm{KS}}(t)\delta t}{\hbar}
\right)
\psi_k(\mathbf{r},t),
\label{eq:propagator}
\end{equation}
which is approximated by a fourth-order Taylor expansion,
\begin{equation}
\psi_k(\mathbf{r},t+\delta t)
\approx
\sum_{n=0}^{4}
\frac{1}{n!}
\left(
-\frac{i\delta t}{\hbar}\hat{H}_{\mathrm{KS}}(t)
\right)^n
\psi_k(\mathbf{r},t).
\label{eq:taylor}
\end{equation}

The propagation is performed for \(N\) time steps until the final simulation
time \(t_{\mathrm{final}}=N\delta t\) is reached. A time step of
\(\delta t=0.001\,\mathrm{fs}\) is used in the simulation. 
Although the fourth-order Taylor propagator is conditionally stable,
it remains accurate for sufficiently small time steps. Its primary
advantage is its simplicity, requiring only repeated applications of
the Hamiltonian to the wavefunction. A review comparing the strengths and limitations of different
time-propagation schemes in TDDFT can be found in~\cite{Castro2004}.

In the real-space TDDFT implementation, the Kohn--Sham orbitals are represented
on a uniform three-dimensional rectangular grid. In the present simulations, we
use a grid spacing of \(0.3~\text{\AA}\) and 100 grid points along each of the
\(x\), \(y\), and \(z\) directions.

To prevent unphysical reflections of the outgoing electron density at the
boundary of the simulation box, a complex absorbing potential (CAP) is applied
near the edges of the grid, following Manolopoulos~\cite{Manolopoulos2002}. The
CAP is written as
\begin{equation}
-iw(x)
=
-i\frac{\hbar^2}{2m}
\left(\frac{2\pi}{\Delta x}\right)^2
f(y),
\label{eq:CAP}
\end{equation}
where \(x_1\) and \(x_2\) define the beginning and end of the absorbing region,
\(\Delta x=x_2-x_1\), \(m\) is the electron mass, and
\begin{equation}
f(y)
=
\frac{4}{c^2}
\left(
\frac{1}{(1+y)^2}
+
\frac{1}{(1-y)^2}
-
2
\right),
\qquad
y=\frac{x-x_1}{\Delta x},
\label{eq:fy}
\end{equation}
with \(c=2.62\). As the molecule is ionized by the laser field, the emitted
electron density is absorbed by the CAP, thereby reducing artificial reflections
from the simulation boundaries.

In each simulation, the ion velocities are initialized according to a Boltzmann
distribution corresponding to 300 K. In typical Coulomb explosion experiments,
molecules are introduced into the chamber through a cold molecular beam, and the
intense laser field rapidly heats them to high temperatures. 
Because the molecular temperature evolves rapidly during Coulomb
explosion, initializing the ionic velocities at 300 K provides a
practical means of sampling diverse fragmentation pathways. If the ion velocities are held fixed, the simulations
produce identical fragmentation outcomes, which limits the statistical
variability of the results. To obtain a broader distribution of fragmentation
behavior, 50 simulations are performed for each molecule. 
This ensemble produces fragmentation events for all three molecules. Owing to limited
computational resources, no additional simulations are performed. Similar
simulation protocols have been used successfully in previous studies of
laser-driven fragmentation dynamics~\cite{Taylor2025,Russakoff2015,Taylor2025PRA,Jiang2025JCP,Jiang2026}.

The molecules considered in this work are \(\mathrm{C_2H_6}\),
\(\mathrm{C_3H_8}\), and \(\mathrm{C_4H_{10}}\). The main results presented here examine the molecular response to  circularly polarized excitation, including both ionization and fragmentation dynamics, for a laser pulse with a width of \(9.18~\mathrm{fs}\) and a wavelength of \(890~\mathrm{nm}\).

\begin{figure}[t]
    \centering
    \begin{subfigure}[t]{0.47\textwidth}
        \centering
        \includegraphics[width=\textwidth]{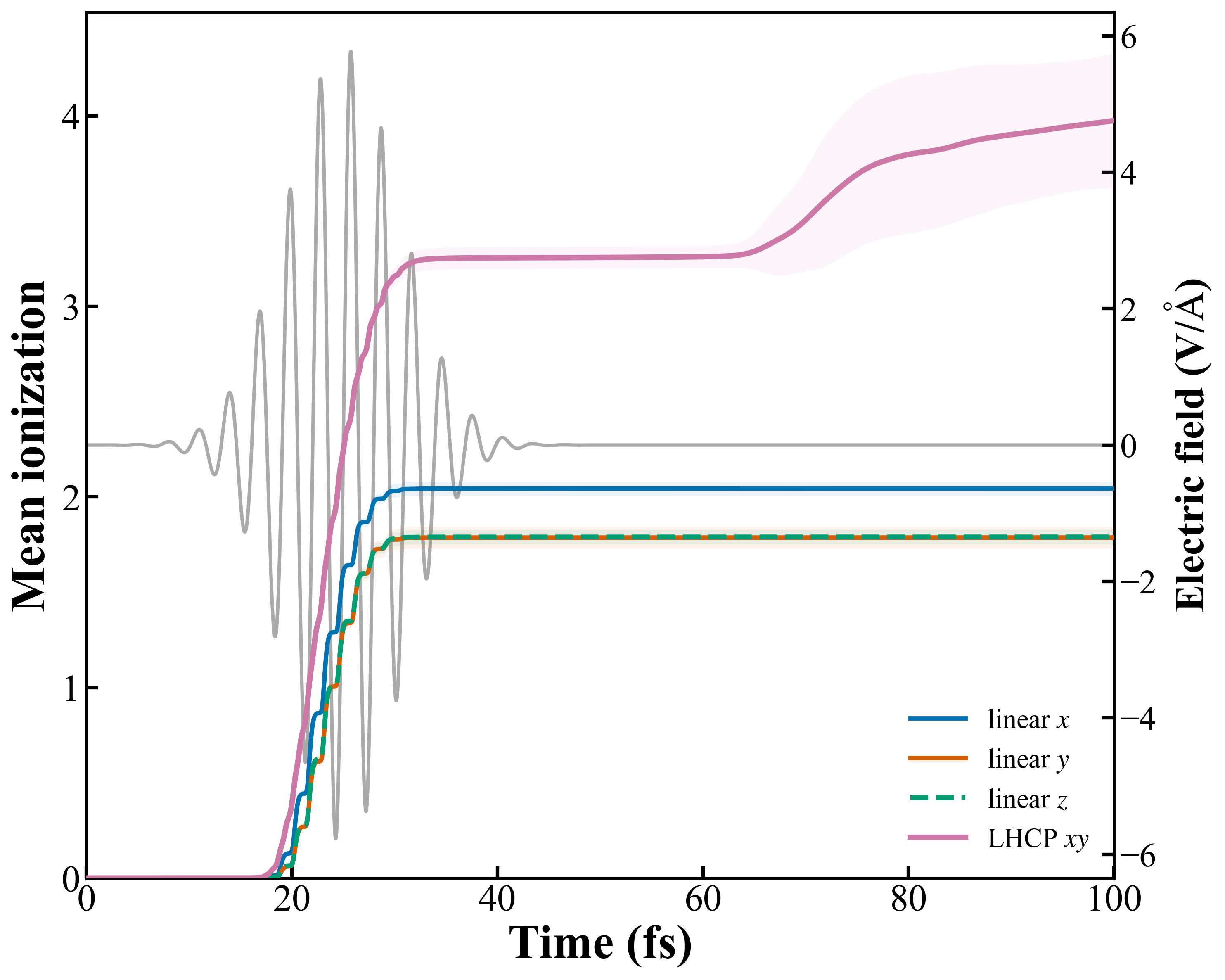}
        \caption{}
        \label{fig:c2h6_ionization_time}
    \end{subfigure}
    \hfill
    \begin{subfigure}[t]{0.47\textwidth}
        \centering
        \includegraphics[width=\textwidth]{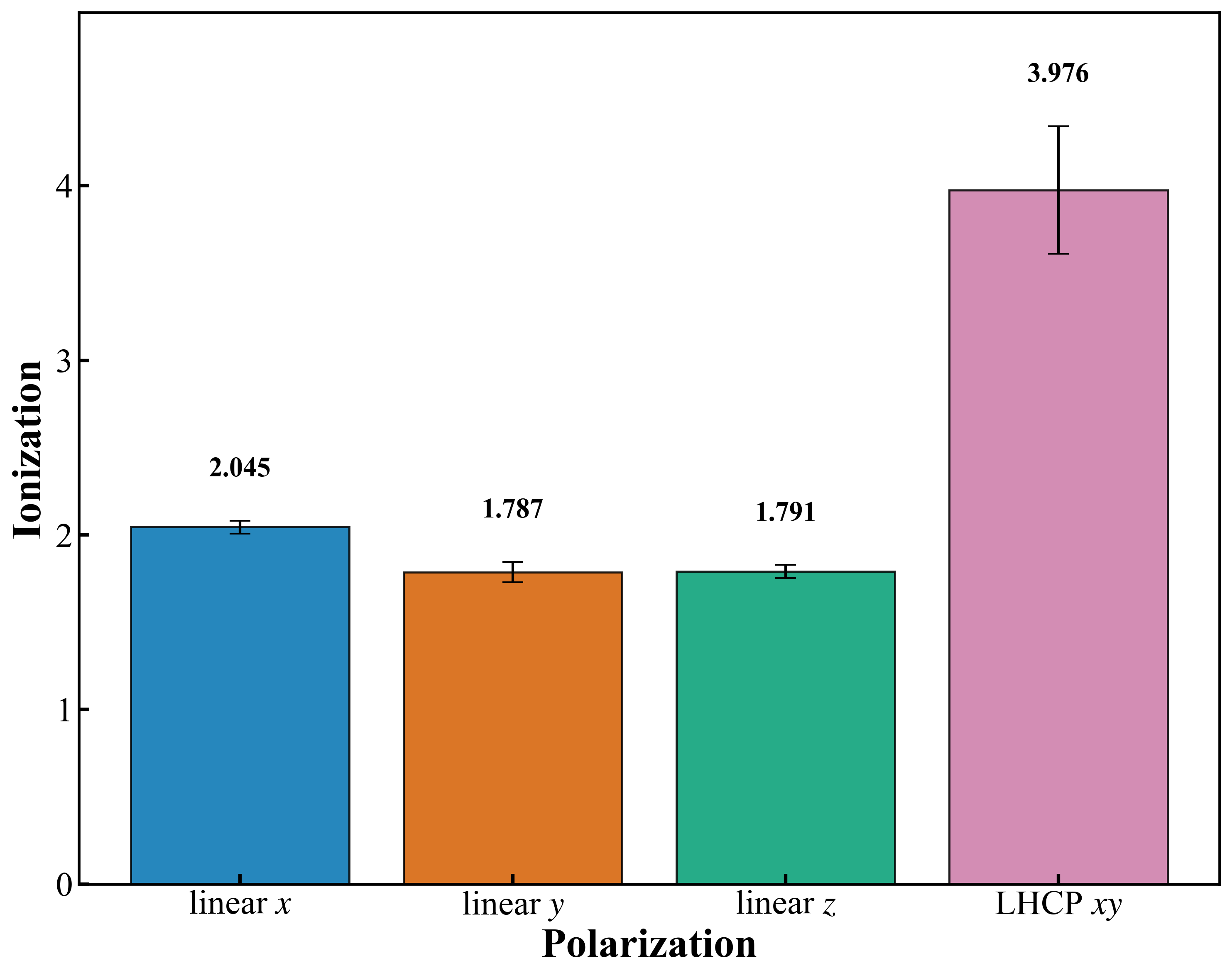}
        \caption{}
        \label{fig:c2h6_ionization_bar}
    \end{subfigure}
\caption{Ionization comparison for \(\mathrm{C_2H_{6}}\) under different laser polarizations.
(a) Time-dependent mean ionization, \(N_e(0)-N_e(t)\), for linear \(x\), \(y\), and \(z\) polarizations and 
Left-Handed Circular Polarization (LHCP)in the \(xy\) plane. Shaded regions indicate trajectory-to-trajectory variation.
(b) Mean ionization at \(100~\mathrm{fs}\), with error bars showing the trajectory-to-trajectory variation.}
    \label{fig:c2h6_ionization}
\end{figure}

\begin{figure}[t]
    \centering
    \begin{subfigure}[t]{0.47\textwidth}
        \centering
        \includegraphics[width=\textwidth]{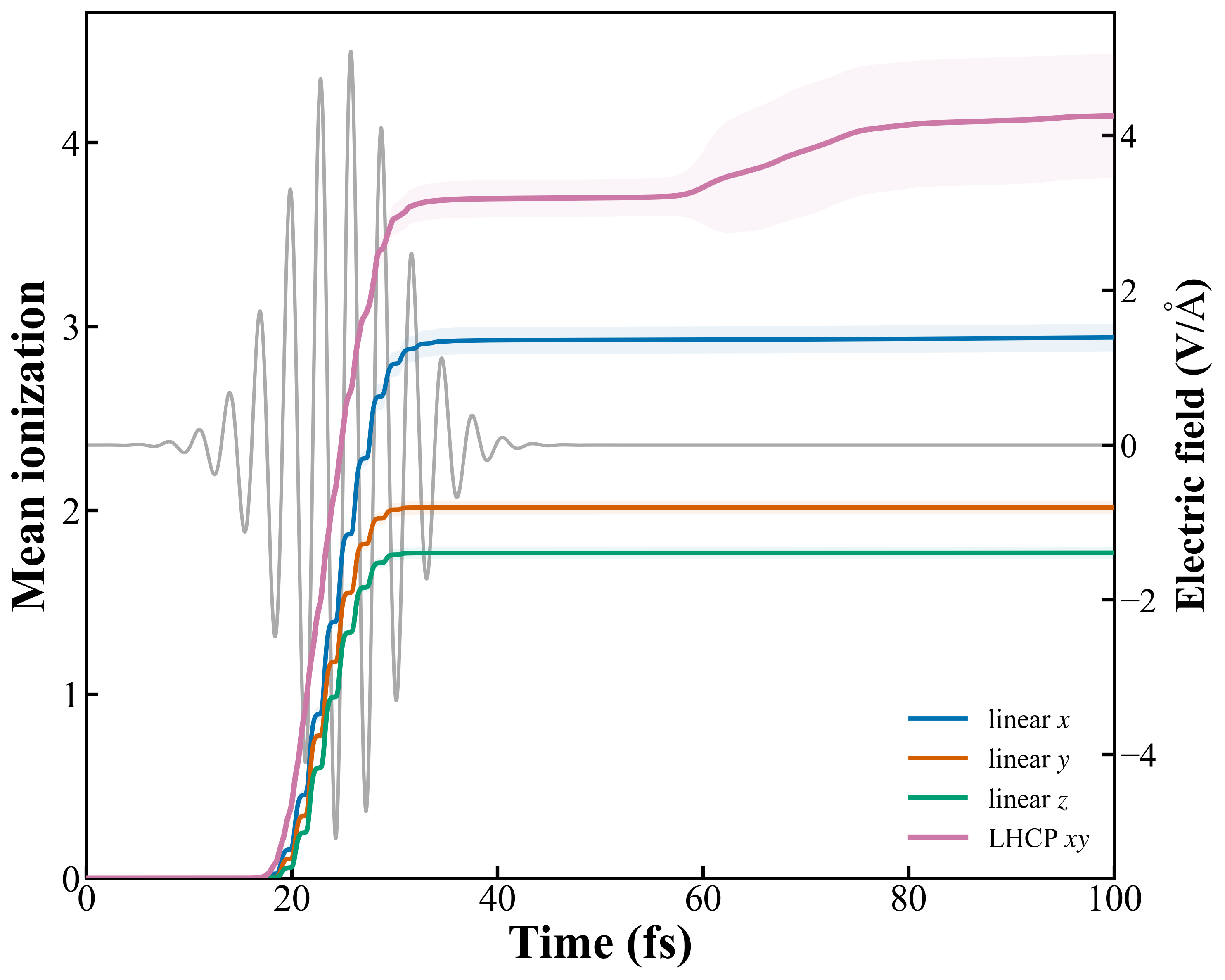}
        \caption{}
        \label{fig:c3h8_ionization_time}
    \end{subfigure}
    \hfill
    \begin{subfigure}[t]{0.47\textwidth}
        \centering
        \includegraphics[width=\textwidth]{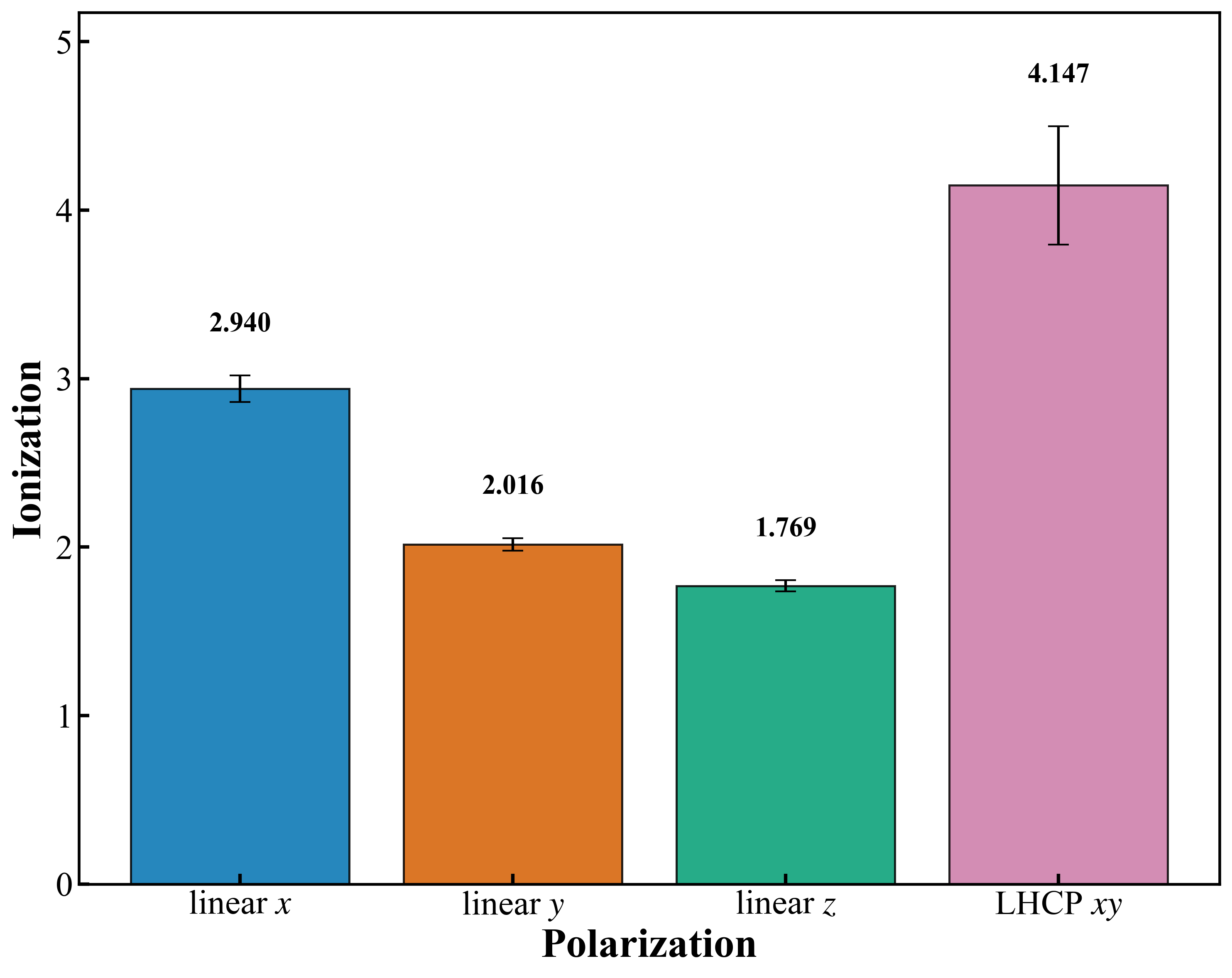}
        \caption{}
        \label{fig:c3h8_ionization_bar}
    \end{subfigure}
\caption{Ionization comparison for \(\mathrm{C_3H_{8}}\) under different laser polarizations.
(a) Time-dependent mean ionization, \(N_e(0)-N_e(t)\), for linear \(x\), \(y\), and \(z\) polarizations and circular polarization in the \(xy\) plane. Shaded regions indicate trajectory-to-trajectory variation.
(b) Mean ionization at \(100~\mathrm{fs}\), with error bars showing the trajectory-to-trajectory variation.}
    \label{fig:c3h8_ionization}
\end{figure}

\begin{figure}[t]
    \centering
    \begin{subfigure}[t]{0.47\textwidth}
        \centering
        \includegraphics[width=\textwidth]{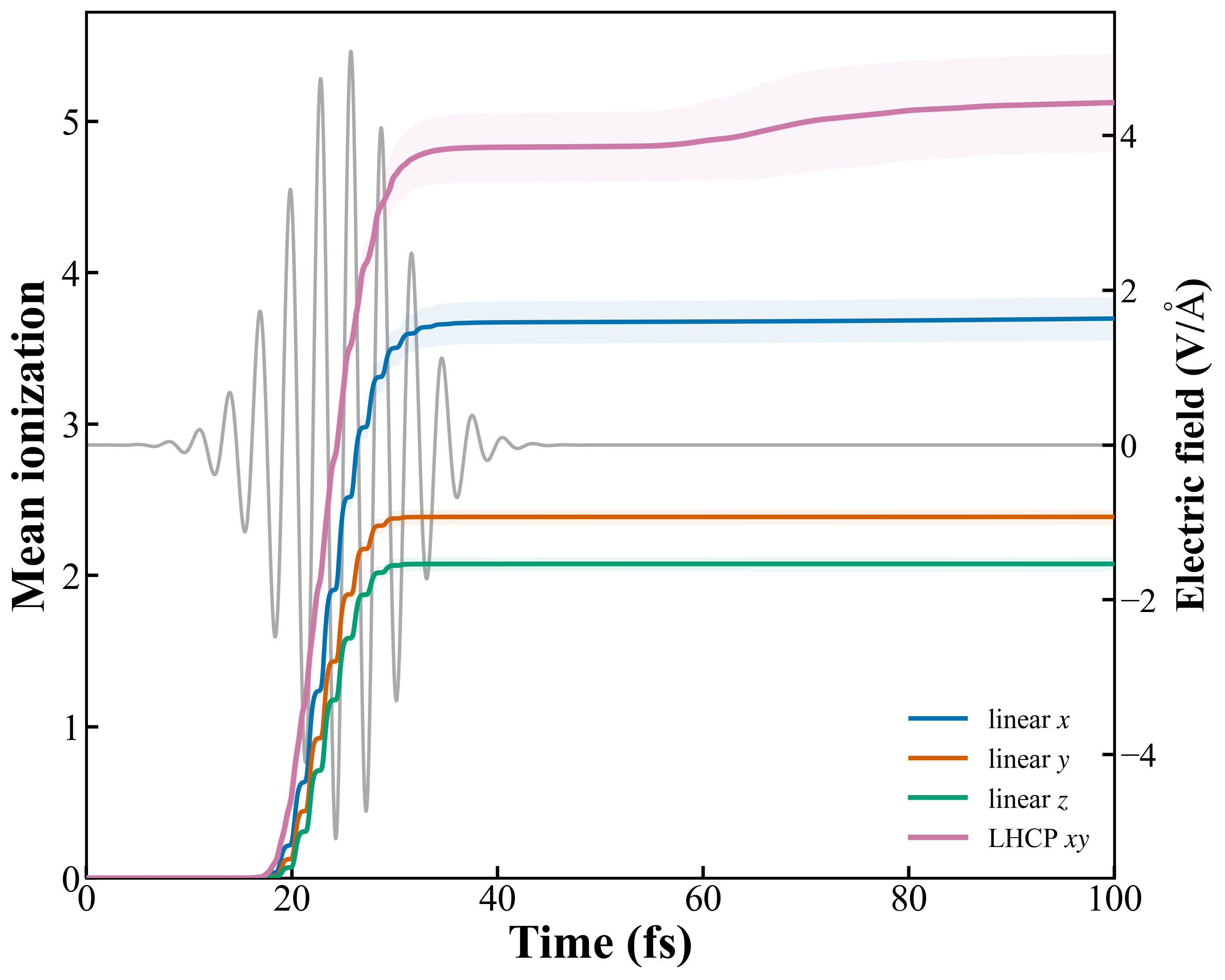}
        \caption{}
        \label{fig:c4h10_ionization_time}
    \end{subfigure}
    \hfill
    \begin{subfigure}[t]{0.47\textwidth}
        \centering
        \includegraphics[width=\textwidth]{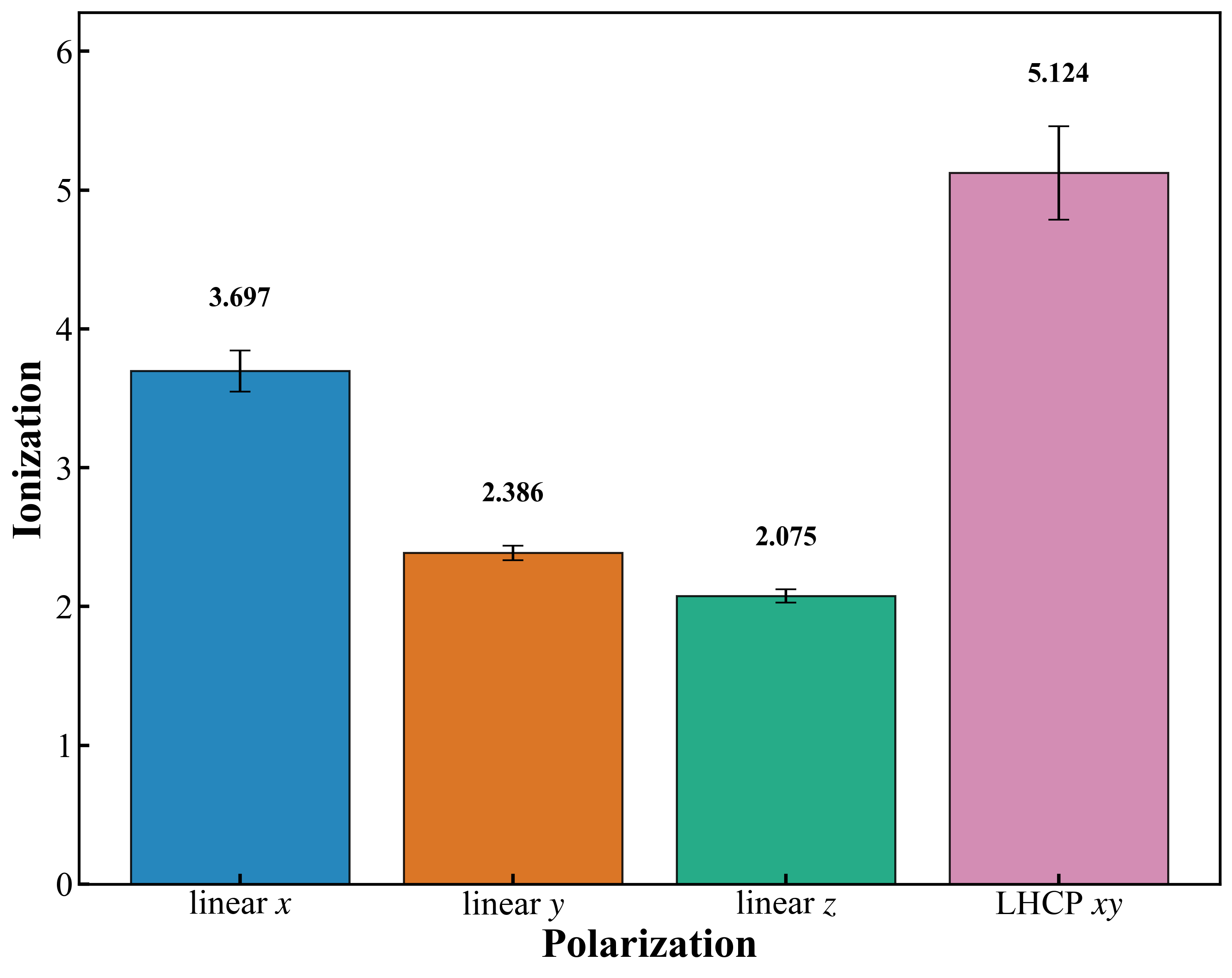}
        \caption{}
        \label{fig:c4h10_ionization_bar}
    \end{subfigure}
\caption{Ionization comparison for \(\mathrm{C_4H_{10}}\) under different laser polarizations.
(a) Time-dependent mean ionization, \(N_e(0)-N_e(t)\), for linear \(x\), \(y\), and \(z\) polarizations and circular polarization in the \(xy\) plane. Shaded regions indicate trajectory-to-trajectory variation.
(b) Mean ionization at \(100~\mathrm{fs}\), with error bars showing the trajectory-to-trajectory variation.}
    \label{fig:c4h10_ionization}
\end{figure}

\section{Results and Discussion}
\subsection{Ionization comparison for different laser polarizations}
To examine how laser polarization influences Coulomb explosion, we first identify the laser intensities that produce clear 
fragmentation for each molecule under circularly polarized excitation. Fragmentation occurs 
at $4.5\times10^{14}~\mathrm{W/cm^2}$ for $\mathrm{C_2H_6}$ and at $3.5\times10^{14}~\mathrm{W/cm^2}$ for 
both $\mathrm{C_3H_8}$ and $\mathrm{C_4H_{10}}$. These intensities are used throughout the ionization, 
fragmentation, and bond-breaking analyses, and the corresponding linearly polarized calculations are 
performed at the same peak intensity so that polarization can be compared directly.
For each molecule and polarization, 15 simulations were performed to compare the ionization dynamics.

Figures~\ref{fig:c2h6_ionization}--\ref{fig:c4h10_ionization} show the time-dependent ionization,
\[
N_e(0)-N_e(t),
\]
for linear polarization along the $x$, $y$, and $z$ axes and for left-handed circular polarization in the $xy$ plane. The ionization dynamics depend strongly on the laser polarization for all three molecules. In each case, circular polarization produces the largest final ionization. For $\mathrm{C_2H_6}$, the mean ionization at $100~\mathrm{fs}$ reaches $3.976$ for circular polarization, compared with $2.045$, $1.787$, and $1.791$ for linear polarization along $x$, $y$, and $z$, respectively. For $\mathrm{C_3H_8}$, circular polarization gives a mean final ionization of $4.147$, while the corresponding linear cases yield $2.940$, $2.016$, and $1.769$. For $\mathrm{C_4H_{10}}$, circular polarization again produces the largest ionization, reaching $5.124$, whereas the linearly polarized fields give $3.697$, $2.386$, and $2.075$.

The curves represent ensemble averages, and the shaded regions indicate trajectory-to-trajectory variation. The polarization dependence emerges during the laser pulse, when the ionization rises rapidly before approaching a plateau. The rotating electric field therefore removes more charge than any fixed linear polarization under the same peak intensity. Among the linear cases, the $x$-polarized field generally produces the largest ionization, while the $y$- and $z$-polarized fields remain lower and are often comparable within the statistical spread.

The later-time increase in the circularly polarized curves should be interpreted with care. Because circular polarization also drives stronger fragmentation under these conditions, part of the continued rise in the computed ionization may arise from atoms or fragments moving toward the complex absorbing potential boundary, where electronic density is removed from the simulation box. This effect is expected to be much smaller for the linearly polarized cases, which produce weaker fragmentation at the same intensity. The early rapid increase therefore provides the clearest measure of direct laser-driven ionization, whereas the later growth in the circularly polarized curves may include contributions from fragment motion and interaction with the absorbing boundary.

Overall, the ionization results identify laser polarization as an important control parameter in the early stages of Coulomb explosion. 
Circular polarization produces the largest ionization for all three alkanes and also leads to fragmentation, 
whereas linear polarization yields smaller ionization and does not drive comparable fragmentation under the same conditions. 
Because the amount of charge removed from the molecule determines the strength of the subsequent Coulomb repulsion 
among ionic fragments, the polarization dependence of ionization provides the basis for understanding the bond-breaking 
and fragmentation dynamics discussed below.

\subsection{Fragmentation products and branching ratios}
Using the molecule-dependent laser intensities identified above, we next analyze the resulting fragmentation products and branching ratios for $\mathrm{C_2H_6}$, $\mathrm{C_3H_8}$, and $\mathrm{C_4H_{10}}$. These intensities correspond to the fragmentation regime for each molecule and are chosen to 
avoid complete dissociation into individual atoms or atomic ions.

\subsubsection{Ethane $\mathrm{C_2H_6}$}

\begin{figure}[t]
    \centering
    \begin{subfigure}[t]{0.4\textwidth}
        \centering
        \includegraphics[width=\linewidth]{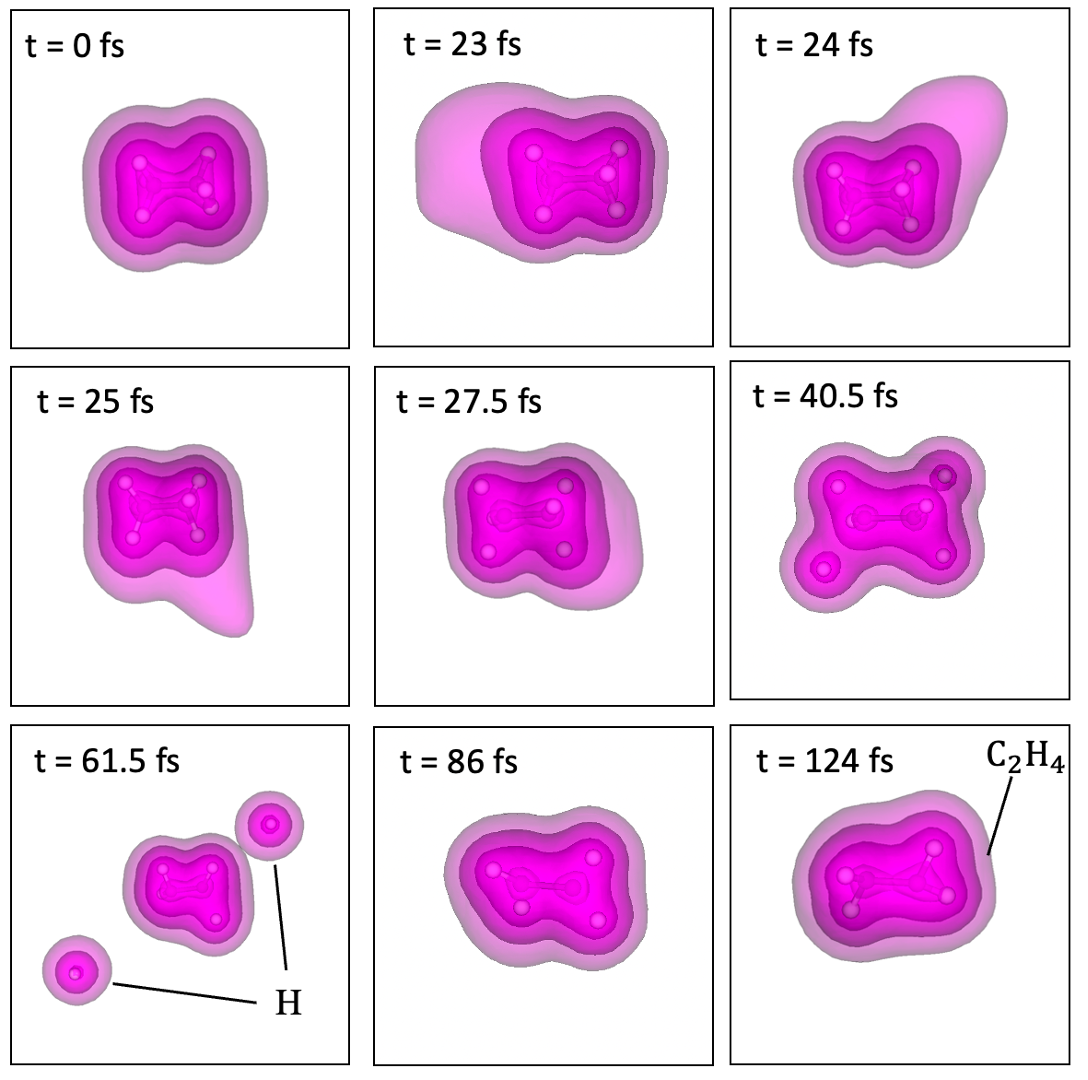}
        \caption{}
        \label{fig:c2h6_snapshots}
    \end{subfigure}
    \begin{subfigure}[t]{0.52\textwidth}
        \centering
        \includegraphics[width=\linewidth]{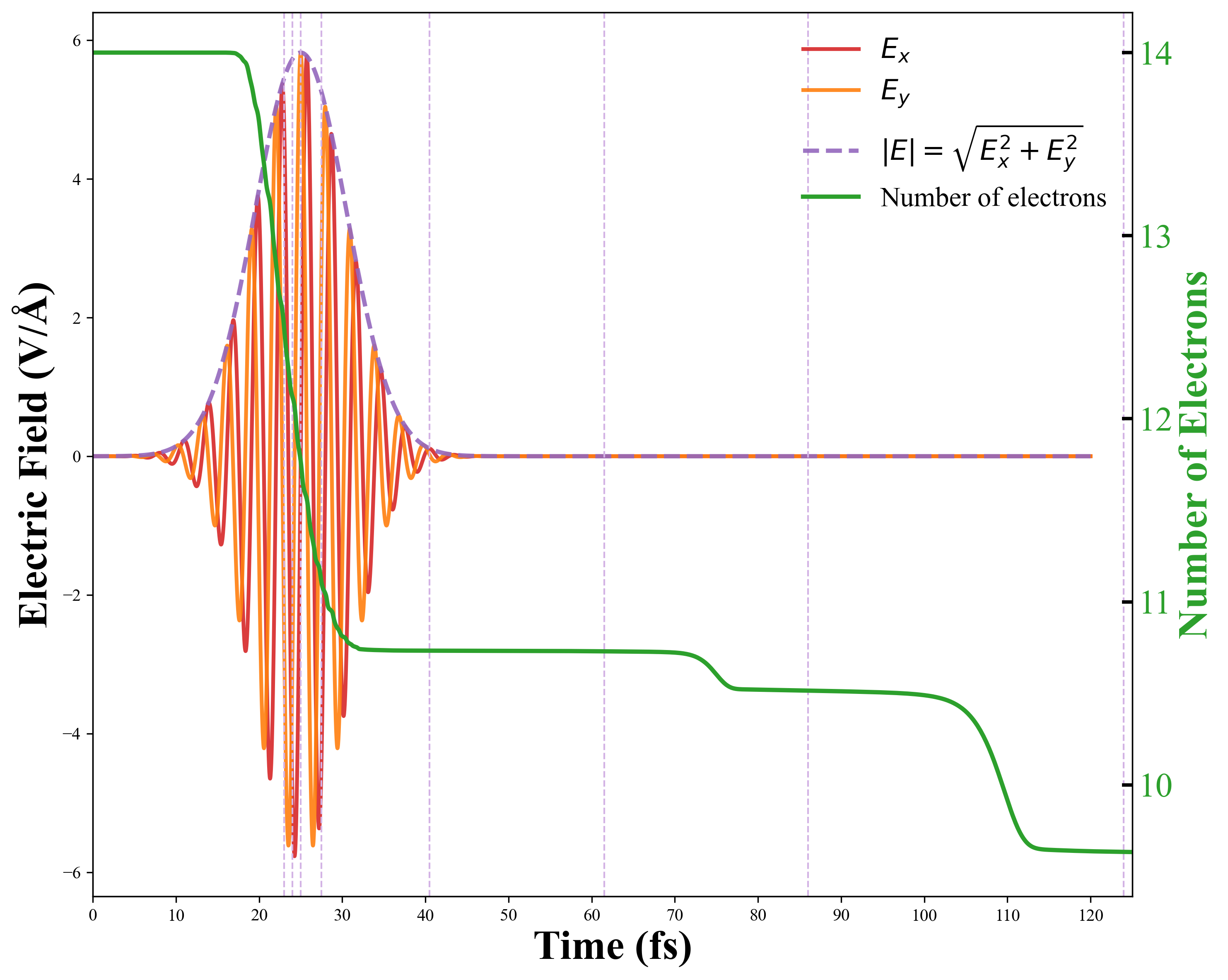}
        \caption{}
        \label{fig:c2h6_electron_light}
    \end{subfigure}
\caption{Coulomb explosion dynamics of a simulation on $\mathrm{C_2H_6}$ yielding $\mathrm{C_2H_4} + 2\mathrm{H}$. (a) Snapshots of Coulomb explosion of $\mathrm{C_2H_6}$ resulting in the formation of $\mathrm{C_2H_4}$ and two $\mathrm{H}$ fragments, carrying charges of $+1.75$, $+0.80$, and $+0.71$, respectively. The 0.001, 0.01, 0.1, and 1 density isosurfaces are shown. (b) Circularly polarized electric field and the corresponding time-dependent number of electrons.}
    \label{fig:c2h6_light_electron}
\end{figure}

\begin{figure}[t]
    \centering
    \begin{subfigure}[t]{0.48\textwidth}
        \centering
        \includegraphics[width=\linewidth]{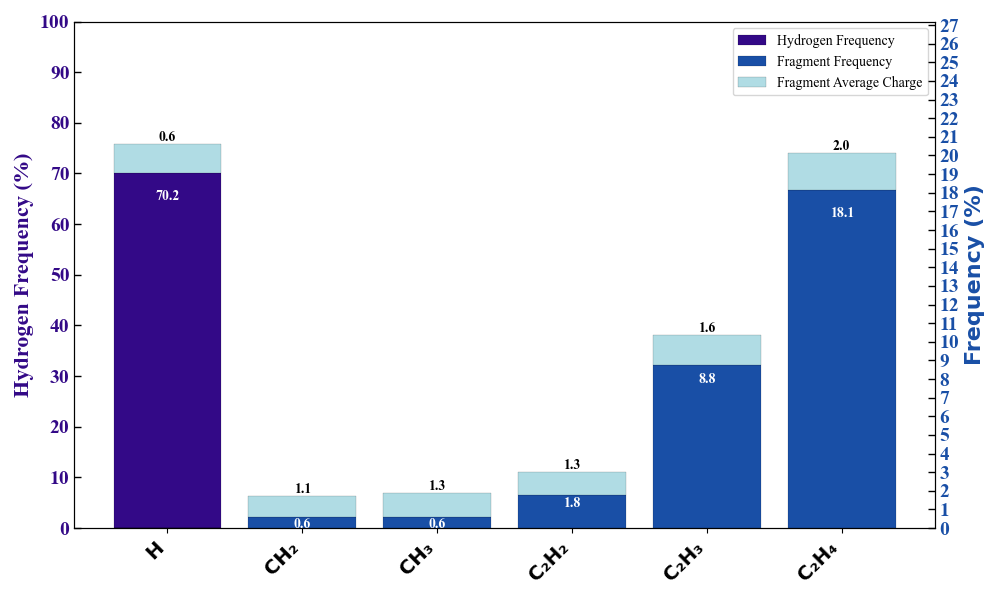}
        \caption{}
        \label{fig:C2H6_frag_charge_averages_two_axes}
    \end{subfigure}
    \begin{subfigure}[t]{0.48\textwidth}
        \centering
        \includegraphics[width=\linewidth]{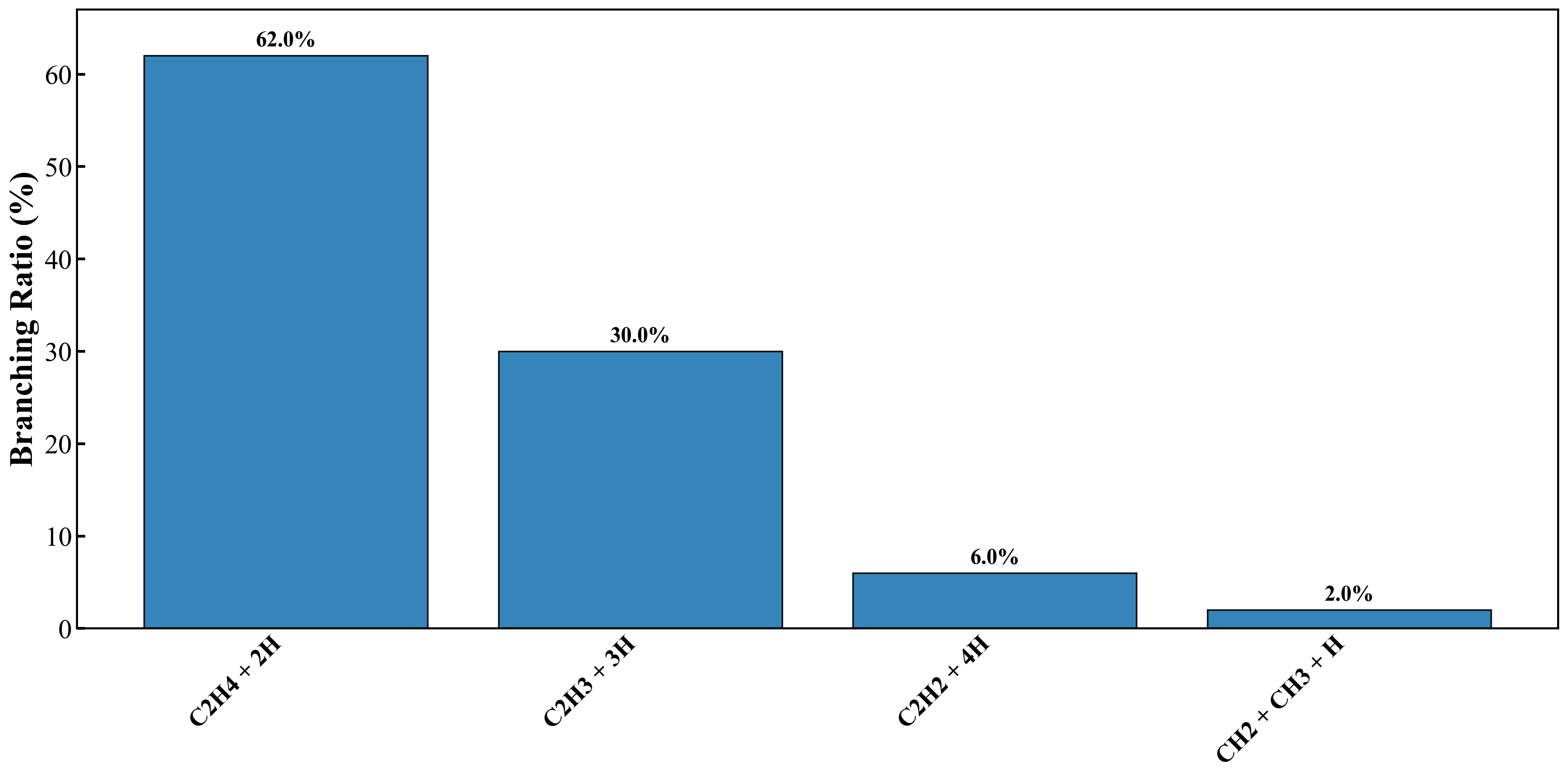}
        \caption{}
        \label{fig:C2H6_channel_branching_ratios}
    \end{subfigure}
    \caption{Statistical analysis of 50 $\mathrm{C_2H_6}$ Coulomb explosion simulations.(a) Fragment occurrence frequencies and hydrogen production frequency. (b) Branching ratios of the fragmentation channels.}
    \label{fig:c2h6_two_pics_horizontal}
\end{figure}

A representative ethane trajectory illustrates the sequence of ionization and nuclear fragmentation under circularly polarized excitation. The electron density is initially localized around the equilibrium geometry and becomes progressively distorted as the pulse rises. Between approximately 23 and 27.5 fs, near the pulse maximum, the density develops pronounced asymmetric extensions away from the molecular framework, indicating that the primary electron-removal stage occurs during the high-field portion of the pulse. The electron-number curve in Fig.~\ref{fig:c2h6_electron_light} confirms this picture: the electron population decreases rapidly during the main pulse, dropping from the initial valence-electron count to approximately 10.7 by 30 fs. After the field vanishes, the electron number remains nearly constant for several tens of femtoseconds before exhibiting additional decreases at later times. These later drops are associated with outgoing electron density reaching the complex absorbing potential rather than with direct laser-driven ionization. The early sharp decrease therefore provides the clearest measure of the field-induced ionization stage.

Nuclear motion proceeds on a significantly longer timescale than electronic ionization. Although the electron density is already strongly distorted near the pulse maximum, the carbon skeleton remains largely intact at early times. By 61.5 fs, hydrogen fragments are visibly detached from the parent hydrocarbon fragment, and at the final time shown the remaining carbon-containing fragment is identified as $\mathrm{C_2H_4}$. This trajectory therefore shows that circularly polarized excitation first removes electronic charge from ethane and subsequently drives Coulomb-induced hydrogen loss while largely preserving the two-carbon backbone.

The fragment statistics show that hydrogen is the most frequently produced fragment across the 50 trajectories, with an occurrence frequency of 70.2\%. Among the carbon-containing fragments, the dominant product is $\mathrm{C_2H_4}$, with a frequency of 18.1\%, followed by $\mathrm{C_2H_3}$ at 8.8\%. The more strongly dehydrogenated fragment $\mathrm{C_2H_2}$ appears with a smaller frequency of 1.8\%, while $\mathrm{CH_2}$ and $\mathrm{CH_3}$ are only weakly produced, each with a frequency of 0.6\%. This behavior is qualitatively consistent with the circularly polarized measurements of Li \textit{et al.}~\cite{Li2020}, where prominent peaks were assigned to dehydrogenated two-carbon fragments such as $\mathrm{C_2H_3^{2+}}$, $\mathrm{C_2H_4^{2+}}$, and $\mathrm{C_2H_5^{2+}}$, together with weaker one-carbon fragments. Thus, both experiment and simulation indicate that ethane fragmentation under circularly polarized excitation is dominated by hydrogen loss while the two-carbon backbone is largely preserved.

The simulated average charges of the main carbon-containing fragments are approximately $+1.8$ for $\mathrm{C_2H_2}$, $+1.6$ for $\mathrm{C_2H_3}$, and $+2.0$ for $\mathrm{C_2H_4}$. These values are broadly consistent with the charge carried by the dehydrogenated ethane fragments observed experimentally, including $\mathrm{C_2H_3^{2+}}$, $\mathrm{C_2H_4^{2+}}$, and $\mathrm{C_2H_5^{2+}}$~\cite{Li2020}. However, the simulated values should be interpreted as density-integrated average fragment charges rather than exact charge states.

To quantify the fragmentation pathways on a trajectory-by-trajectory basis, we define the branching ratio of a channel as the fraction of simulated trajectories that produce a given final fragment combination. For a channel $i$, the branching ratio is
\[
B_i=\frac{N_i}{N_{\mathrm{tot}}}\times100\%,
\]
where $N_i$ is the number of trajectories ending in channel $i$ and $N_{\mathrm{tot}}$ is the total number of trajectories. In the present ethane calculations, $N_{\mathrm{tot}}=50$.

The branching ratios in Fig.~\ref{fig:c2h6_two_pics_horizontal} provide a trajectory-level view of the same trend. The dominant pathway is $\mathrm{C_2H_4}+2\mathrm{H}$, which accounts for 62.0\% of the trajectories, followed by $\mathrm{C_2H_3}+3\mathrm{H}$ at 30.0\%. The more strongly dehydrogenated channel $\mathrm{C_2H_2}+4\mathrm{H}$ contributes 6.0\%, whereas the backbone-cleavage channel $\mathrm{CH_2}+\mathrm{CH_3}+\mathrm{H}$ is rare, with a branching ratio of only 2.0\%.

\subsubsection{Propane $\mathrm{C_3H_8}$}

\begin{figure}[t]
    \centering
    \begin{subfigure}[t]{0.5\textwidth}
        \centering
        \includegraphics[width=\linewidth]{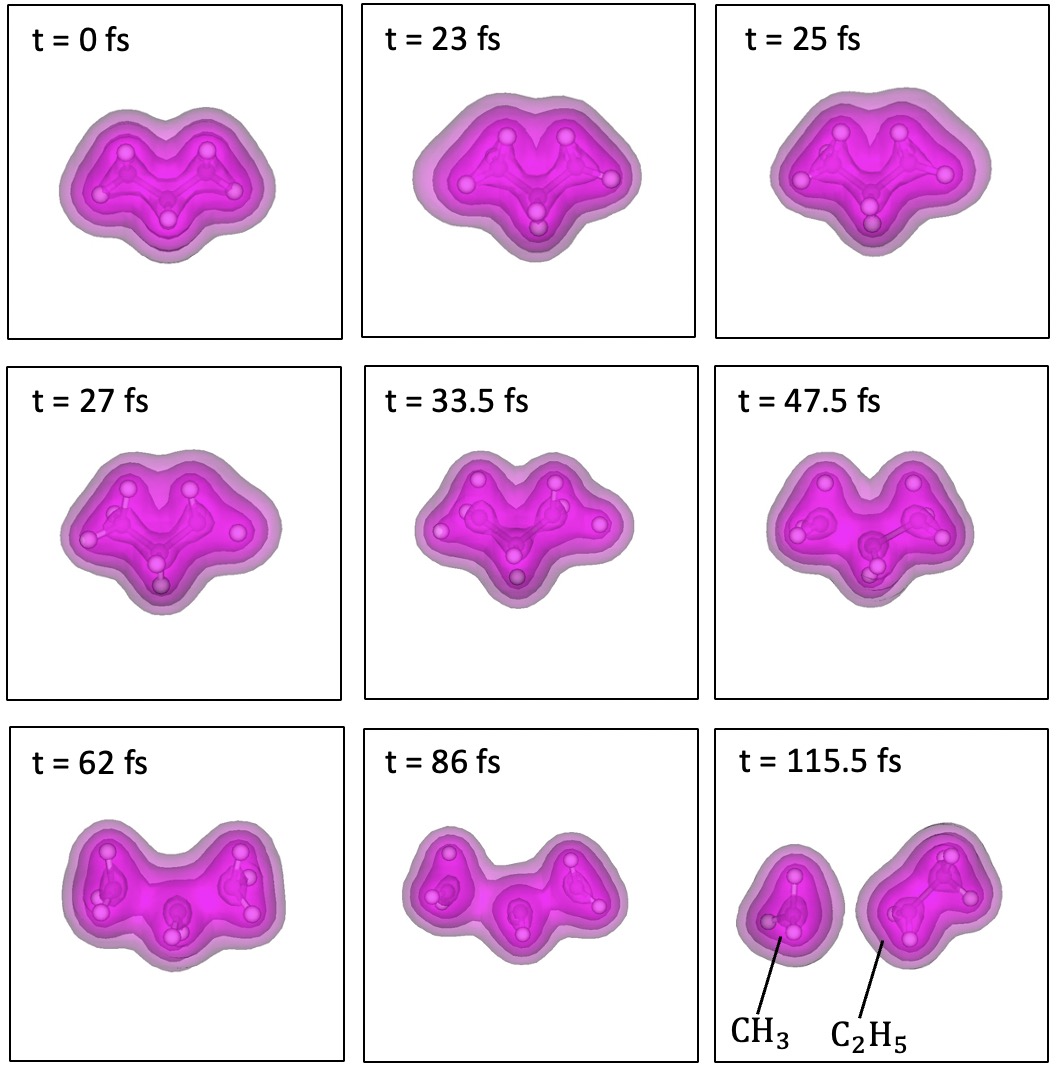}
        \caption{}
        \label{fig:c3h8_snapshot}
    \end{subfigure}
    \begin{subfigure}[t]{0.5\textwidth}
        \centering
        \includegraphics[width=\linewidth]{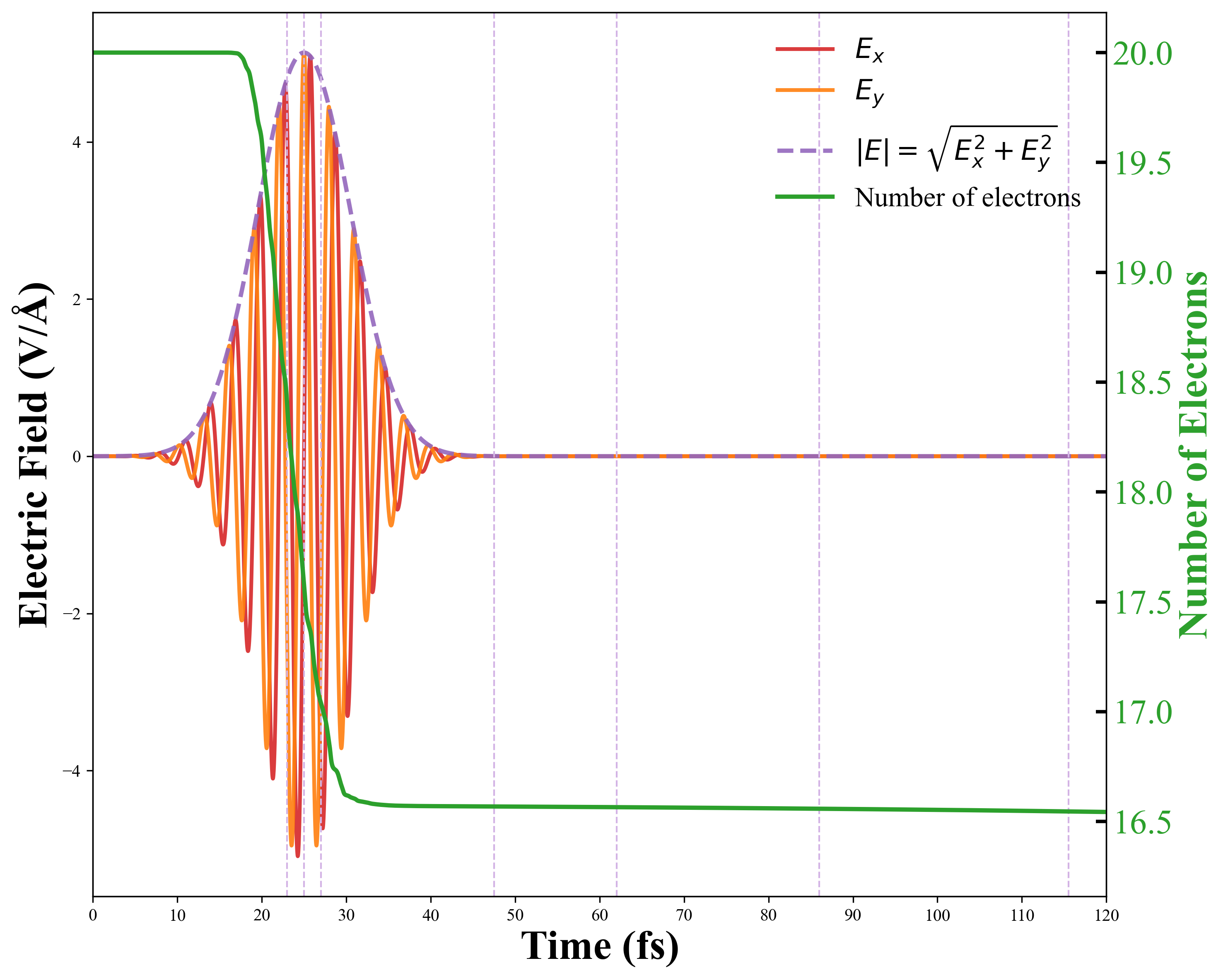}
        \caption{}
        \label{fig:c3h8_electron_light}
    \end{subfigure}
    \caption{Coulomb explosion dynamics of a simulation on $\mathrm{C_3H_8}$ yielding $\mathrm{CH_3} + \mathrm{C_2H_5}$. (a) Snapshots of Coulomb explosion of $\mathrm{C_3H_8}$ resulting in the formation of $\mathrm{CH_3}$ and $\mathrm{C_2H_5}$, carrying charges of $+1.39$ and $+2.13$, respectively. The 0.001, 0.01, 0.1, and 1 density isosurfaces are shown. (b) Circularly polarized electric field and the corresponding time-dependent number of electrons.}
    \label{fig:c3h8_light_electron}
\end{figure}

\begin{figure}[t]
    \centering
    \begin{subfigure}[t]{0.48\textwidth}
        \centering
        \includegraphics[width=\linewidth]{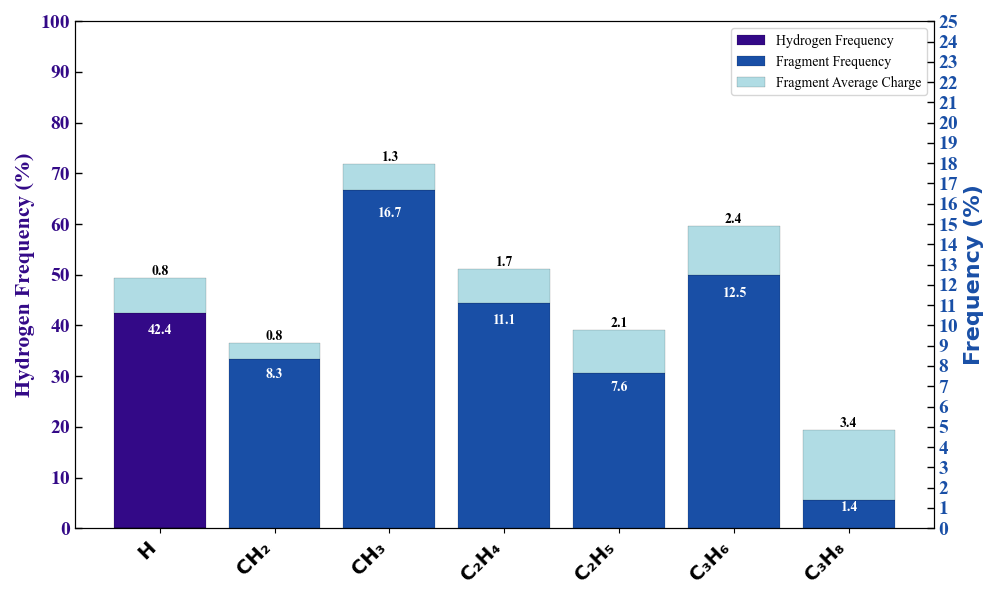}
        \caption{}
        \label{fig:C3H8_frag_charge_averages_two_axes}
    \end{subfigure}
    \begin{subfigure}[t]{0.48\textwidth}
        \centering
        \includegraphics[width=\linewidth]{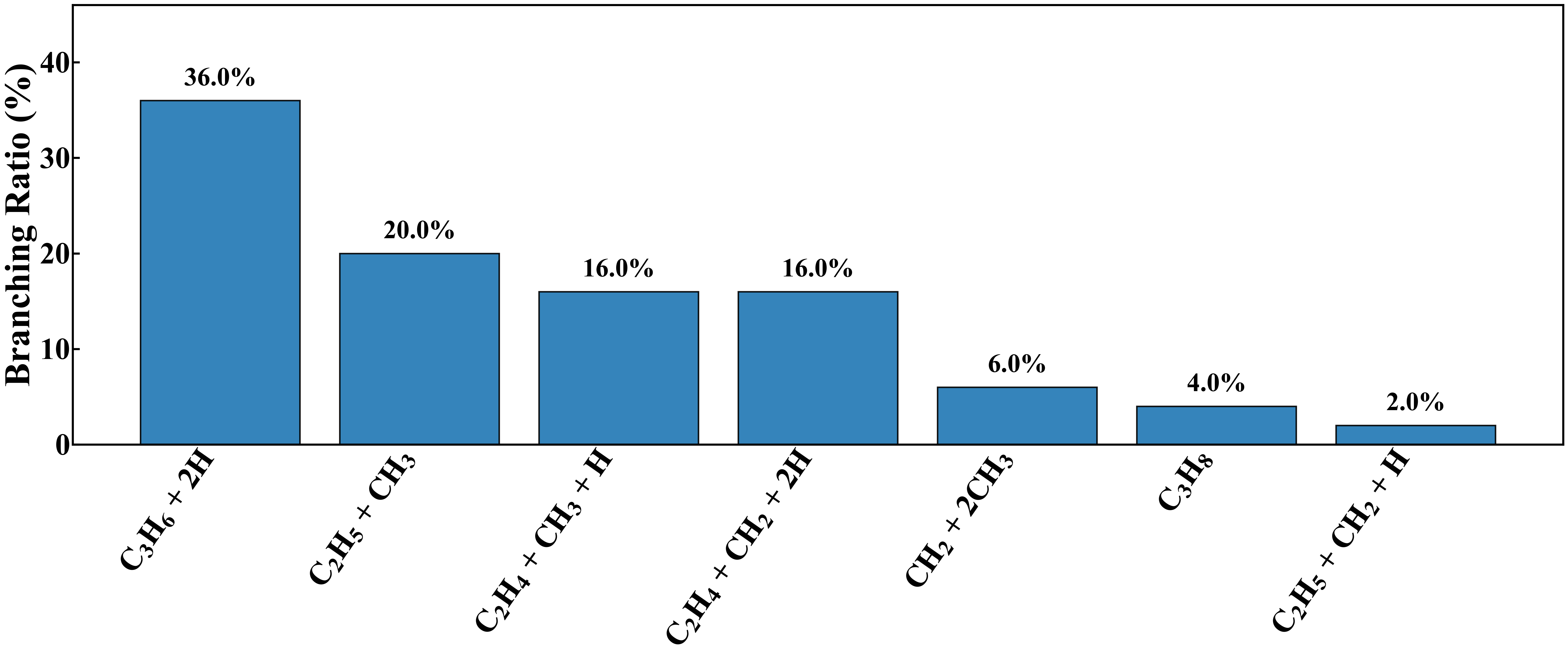}
        \caption{}
        \label{fig:C3H8_channel_branching_ratios}
    \end{subfigure}
    \caption{Statistical analysis of 50 $\mathrm{C_3H_8}$ Coulomb explosion. (a) Fragment occurrence frequencies and hydrogen production frequency. (b) Branching ratios of the fragmentation channels.}
    \label{fig:c3h8_statistics}
\end{figure}

Propane shows the same general sequence of rapid laser-driven ionization followed by delayed nuclear breakup, but its fragmentation landscape is broader than that of ethane. During the pulse, the rotating electric field strongly distorts the electron density and drives charge removal near the field maximum. Fragmentation develops only after sufficient positive charge has accumulated on the molecular framework. In the representative trajectory shown in Fig.~\ref{fig:c3h8_light_electron}, the carbon backbone eventually separates, and by 115.5 fs the system has dissociated mainly into $\mathrm{CH_3}$ and $\mathrm{C_2H_5}$ fragments.

The fragment statistics show that atomic hydrogen remains the most frequently produced fragment, with an occurrence frequency of 42.4\%. Among the carbon-containing products, the dominant fragments are $\mathrm{CH_3}$ (16.7\%), $\mathrm{C_3H_6}$ (12.5\%), and $\mathrm{C_2H_4}$ (11.1\%), with smaller contributions from $\mathrm{CH_2}$ (8.3\%) and $\mathrm{C_2H_5}$ (7.6\%). The small population of intact $\mathrm{C_3H_8}$ fragments (1.4\%) indicates that most trajectories undergo either dehydrogenation or partial backbone cleavage. The larger hydrocarbon fragments also carry higher average charges, increasing from about $+1.3$ for $\mathrm{CH_3}$ to $+2.4$ for $\mathrm{C_3H_6}$, suggesting that the residual positive charge is mainly retained by the larger carbon-containing fragments.

The branching ratios in Fig.~\ref{fig:c3h8_statistics} further clarify the dominant dissociation pathways. The largest channel is $\mathrm{C_3H_6}+2\mathrm{H}$, with a branching ratio of 36\%, showing that dehydrogenation with preservation of the three-carbon backbone is the most probable outcome. Backbone cleavage is also significant, as seen from the $\mathrm{C_2H_5}+\mathrm{CH_3}$ channel (20\%) and the $\mathrm{C_2H_4}+\mathrm{CH_3}+\mathrm{H}$ and $\mathrm{C_2H_4}+\mathrm{CH_2}+2\mathrm{H}$ channels, each contributing 16\%. Overall, propane fragmentation under circularly polarized excitation is characterized by both hydrogen removal and moderate C--C bond cleavage, while complete breakup into several small carbon fragments remains comparatively unlikely.

\subsubsection{\texorpdfstring{$n$-Butane $\mathrm{C_4H_{10}}$}{n-Butane C4H10}}

\begin{figure}[t]
    \centering
    \begin{subfigure}[t]{0.5\textwidth}
        \centering
        \includegraphics[width=\linewidth]{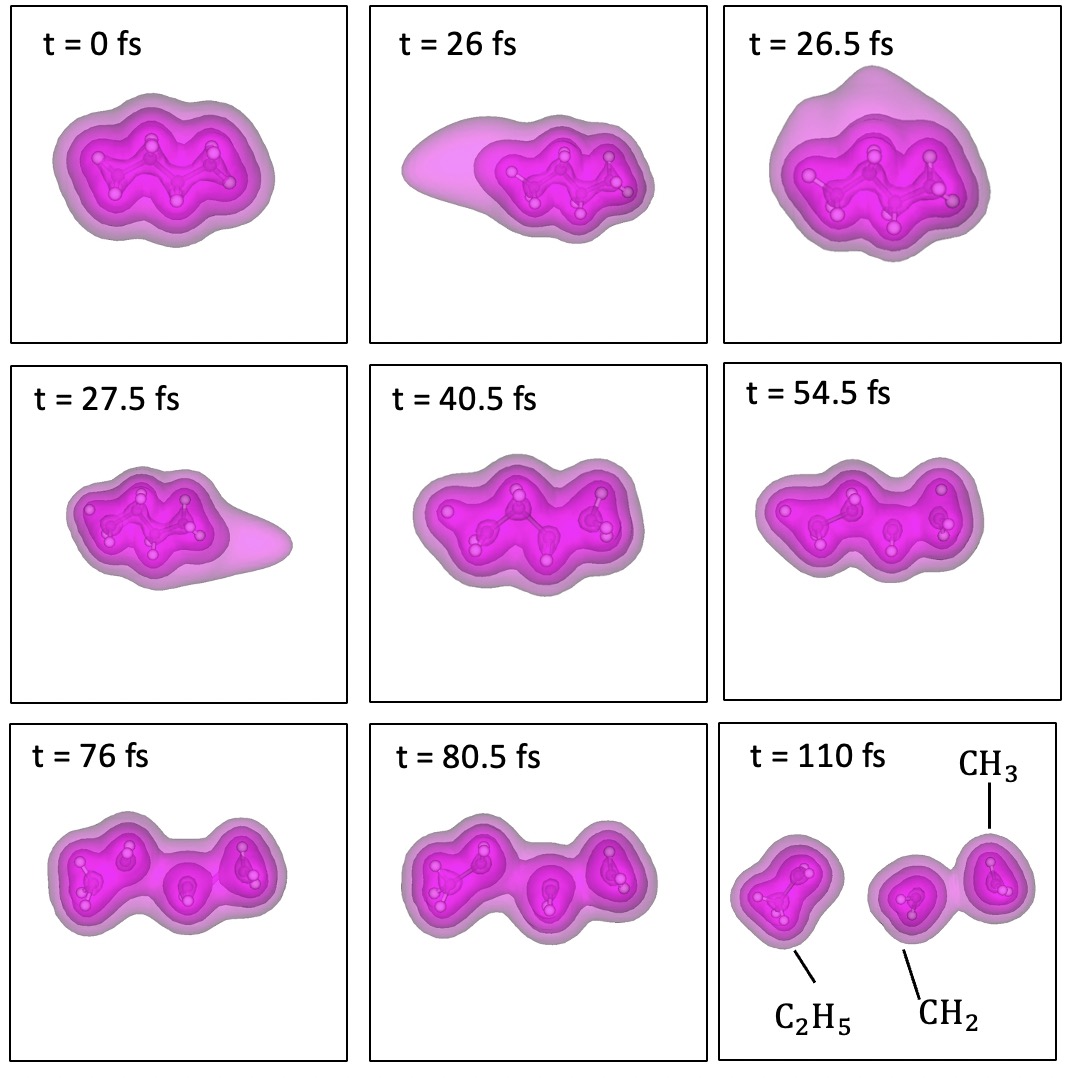}
        \caption{}
        \label{fig:c4h10_snapshot}
    \end{subfigure}
    \begin{subfigure}[t]{0.5\textwidth}
        \centering
        \includegraphics[width=\linewidth]{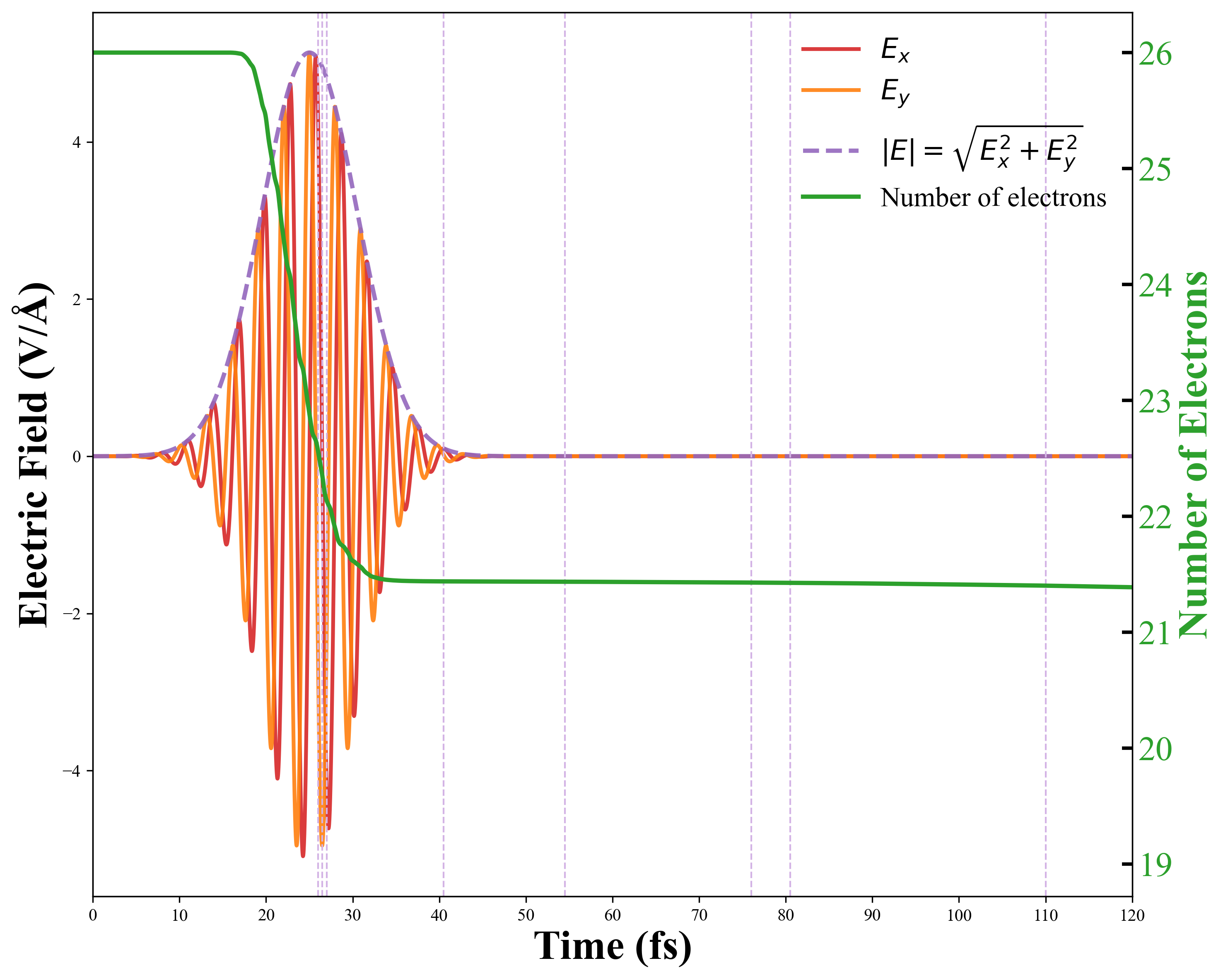}
        \caption{}
        \label{fig:c4h10_electron_light}
    \end{subfigure}
    \caption{Coulomb explosion dynamics of a simulation on $\mathrm{C_4H_{10}}$ yielding $\mathrm{C_2H_5} + \mathrm{CH_2} + \mathrm{CH_3}$. (a) Snapshots of Coulomb explosion of $\mathrm{C_4H_{10}}$ showing the formation of one $\mathrm{C_2H_5}$, $\mathrm{CH_2}$, and $\mathrm{CH_3}$, which carry charges of $+2.23$, $+1.04$, and $+1.35$, respectively. The 0.001, 0.01, 0.1, and 1 electron-density isosurfaces are displayed. (b) Circularly polarized electric field and the corresponding time-dependent number of electrons.}
    \label{fig:c4h10_light_electron}
\end{figure}

\begin{figure}[h]
    \centering
    \begin{subfigure}[t]{0.48\textwidth}
        \centering
        \includegraphics[width=\linewidth]{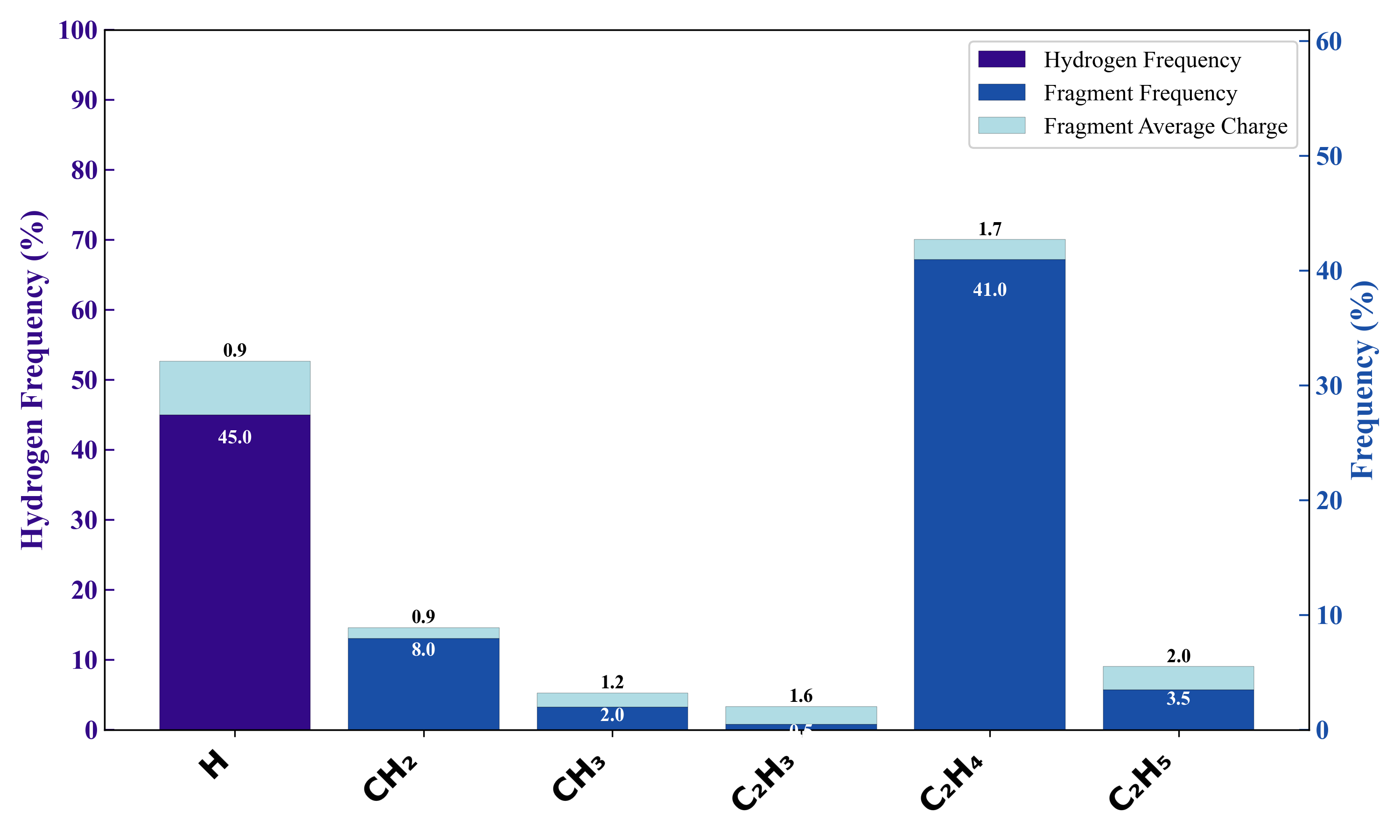}
        \caption{}
        \label{fig:C4H10_frag_charge_averages_two_axes}
    \end{subfigure}
    \hfill
    \begin{subfigure}[t]{0.5\textwidth}
        \centering
        \includegraphics[width=\linewidth]{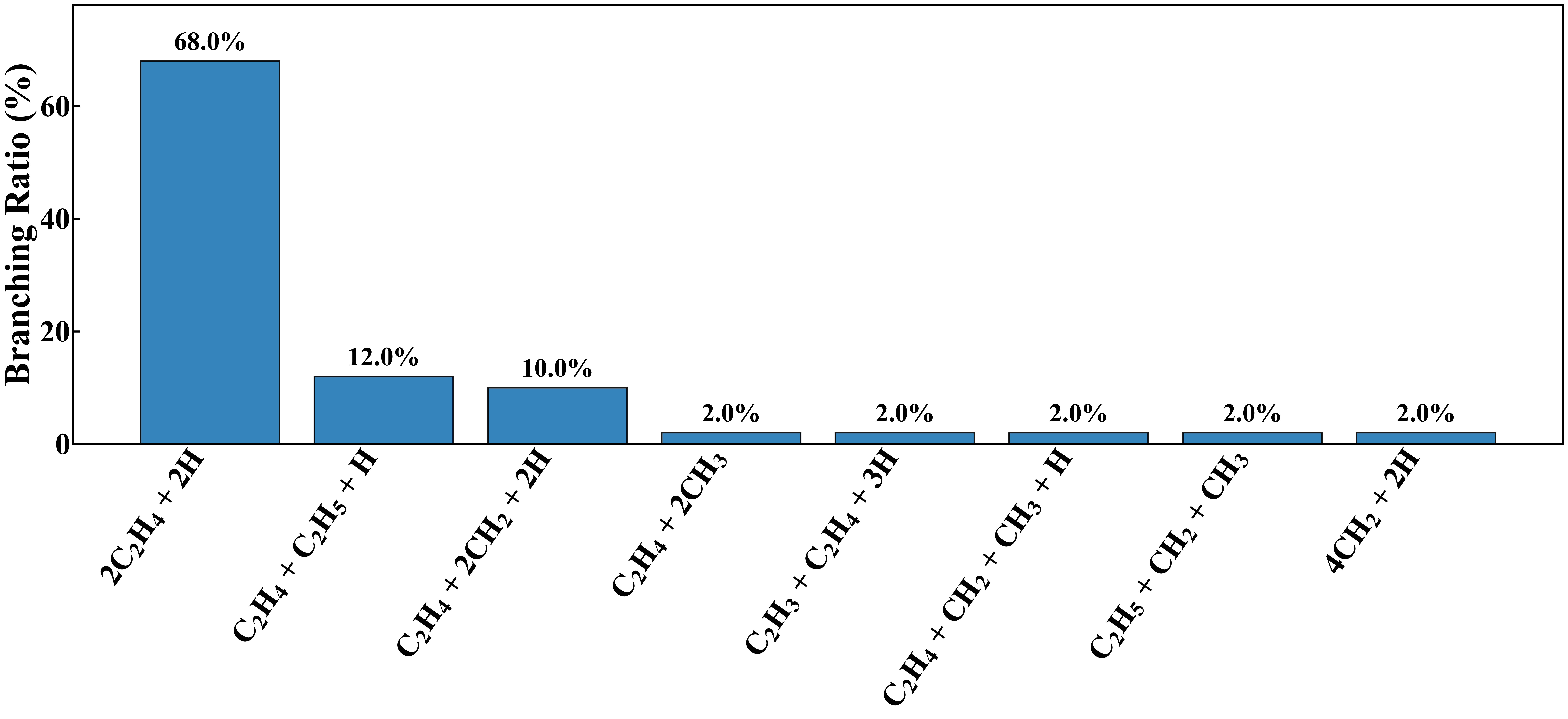}
        \caption{}
        \label{fig:C4H10_channel_branching_ratios}
    \end{subfigure}
    \caption{Statistical analysis of 50 $\mathrm{C_4H_{10}}$ Coulomb explosion. (a) Fragment occurrence frequencies and hydrogen production frequency. (b) Branching ratios of the fragmentation channels.}
    \label{fig:c4h10_statistics}
\end{figure}

Butane follows the same overall pattern of rapid ionization and delayed nuclear breakup, but its larger carbon skeleton opens additional fragmentation pathways. The electron density is initially localized around the intact molecular structure. Near the pulse maximum, around 26--27.5 fs, the density becomes strongly distorted and develops asymmetric extensions, indicating substantial ionization driven by the rotating field. The corresponding electron-number curve shows a rapid decrease during the main pulse, followed by a nearly constant plateau after the field vanishes. Clear structural separation appears later, and by 110 fs the molecule has dissociated into $\mathrm{C_2H_5}$, $\mathrm{CH_2}$, and $\mathrm{CH_3}$ fragments.

The statistical fragment distribution in Fig.~\ref{fig:c4h10_statistics} shows that H and $\mathrm{C_2H_4}$ are the dominant products of butane fragmentation under circularly polarized excitation, with occurrence frequencies of 45.0\% and 41.0\%, respectively. Much smaller contributions arise from $\mathrm{CH_2}$ (8.0\%), $\mathrm{C_2H_5}$ (3.5\%), $\mathrm{CH_3}$ (2.0\%), and $\mathrm{C_2H_3}$ (1.0\%). The average fragment charge is approximately $0.9$ for both $\mathrm{H}$ and $\mathrm{CH_2}$ and increases to $1.2$ for $\mathrm{CH_3}$, $1.6$ for $\mathrm{C_2H_3}$, $1.7$ for $\mathrm{C_2H_4}$, and $2.0$ for $\mathrm{C_2H_5}$. These results indicate that butane fragmentation is dominated by hydrogen loss together with the formation of relatively stable two-carbon hydrocarbon fragments, especially $\mathrm{C_2H_4}$.

The branching-ratio distribution in Fig.~\ref{fig:C4H10_channel_branching_ratios} further shows that the dominant channel is $2\mathrm{C_2H_4}+2\mathrm{H}$, with a branching ratio of 68\%. This indicates that the most probable dissociation pathway involves dehydrogenation and cleavage of the carbon backbone into two ethylene-like fragments. The next most important channels are $\mathrm{C_2H_4}+\mathrm{C_2H_5}+\mathrm{H}$ (12\%) and $\mathrm{C_2H_4}+2\mathrm{CH_2}+2\mathrm{H}$ (10\%). All remaining channels contribute only 2\% each, including pathways involving $\mathrm{CH_3}$, $\mathrm{C_2H_3}$, or multiple $\mathrm{CH_2}$ fragments. Overall, the Coulomb explosion of butane under circular polarization is strongly dominated by partial backbone cleavage into relatively large hydrocarbon fragments, while complete breakup into several small carbon-containing fragments is rare.

Taken together, these results show that hydrogen loss is the dominant fragmentation response of all three alkanes under circularly polarized excitation, while the carbon-containing products exhibit a clear dependence on molecular size. For $\mathrm{C_2H_6}$, fragmentation is dominated by sequential dehydrogenation, producing primarily $\mathrm{C_2H_4}$ and $\mathrm{C_2H_3}$ fragments while largely preserving the two-carbon backbone. In $\mathrm{C_3H_8}$, dehydrogenation remains important, with $\mathrm{C_3H_6}+2\mathrm{H}$ forming the largest channel, but C--C bond cleavage also contributes substantially through pathways such as $\mathrm{C_2H_5}+\mathrm{CH_3}$ and $\mathrm{C_2H_4}+\mathrm{CH_3}+\mathrm{H}$. For $\mathrm{C_4H_{10}}$, the dominant channel shifts toward backbone cleavage into two-carbon fragments, most notably $2\mathrm{C_2H_4}+2\mathrm{H}$. Thus, increasing molecular size broadens the accessible fragmentation pathways and enhances the role of C--C cleavage. However, even for the larger alkanes, the Coulomb explosion does not proceed primarily through complete atomization; instead, it favors early hydrogen removal and the formation of relatively stable hydrocarbon fragments.

\subsection{Bond-Breaking Dynamics}
\begin{figure}[h]
    \centering
    \includegraphics[width=1\linewidth]{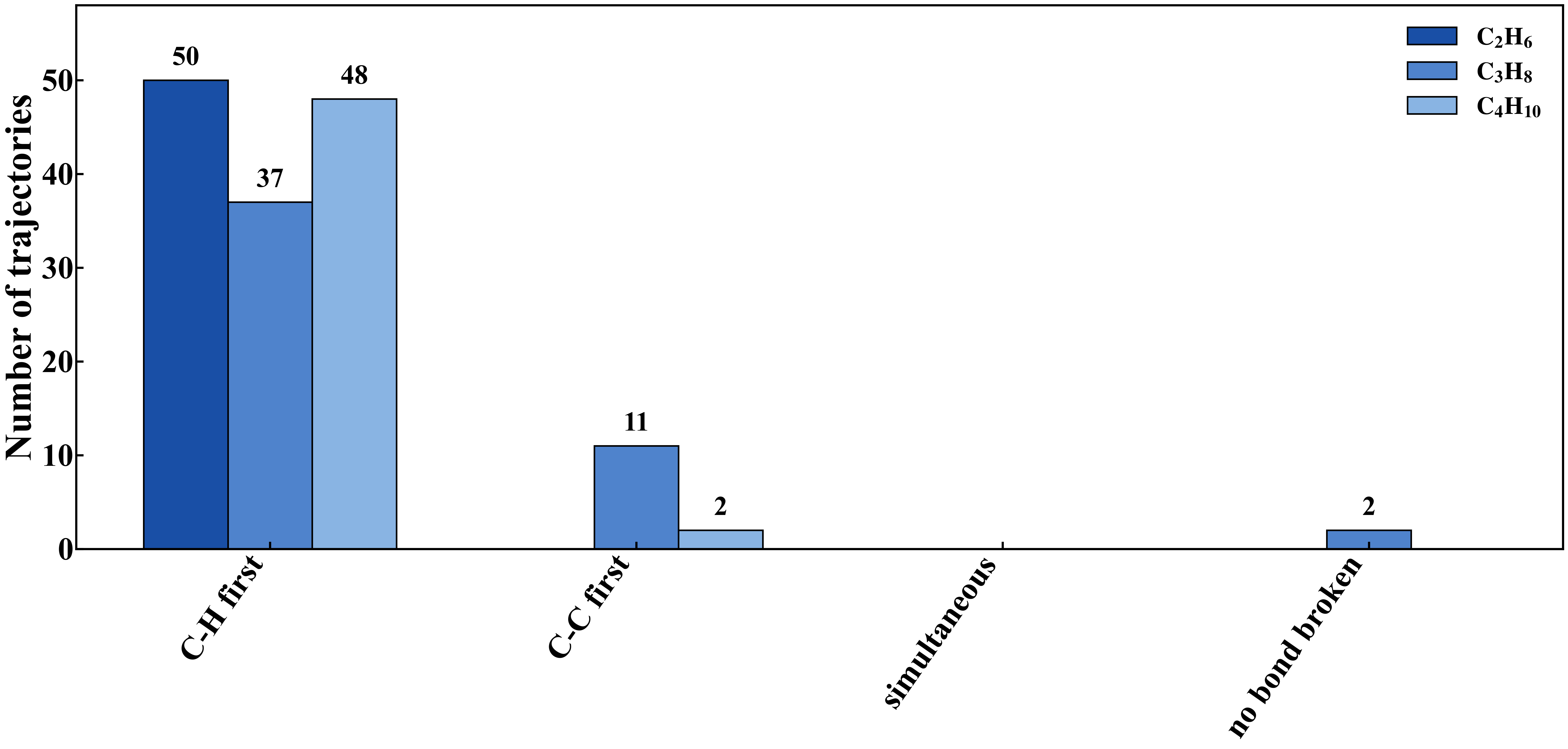}
    \caption{Frequency of the earliest bond-breaking event for $\mathrm{C_2H_6}$, $\mathrm{C_3H_8}$, and $\mathrm{C_4H_{10}}$.}
    \label{fig:combined_bar_which_breaks_first}
\end{figure}

The fragment statistics capture the final outcomes of Coulomb explosion but do not reveal the order in which bonds fail. To address that point, we analyze the identity and timing of the first bond-breaking event over 50 trajectories for each molecule. A C--H bond is defined as broken when its bond length exceeds $2~\text{\AA}$, while a C--C bond is defined as broken when its 
bond length exceeds $2.2~\text{\AA}$. Figure~\ref{fig:combined_bar_which_breaks_first} shows that C--H cleavage is the dominant first bond-breaking channel for all three molecules. In $\mathrm{C_2H_6}$, all 50 trajectories exhibit C--H-first dissociation. In $\mathrm{C_3H_8}$, 37 trajectories are C--H-first, 11 are C--C-first, and 2 show no bond breaking within the analyzed time window. In $\mathrm{C_4H_{10}}$, C--H-first dissociation again dominates, occurring in 48 trajectories, while only 2 trajectories are C--C-first. These results indicate that early hydrogen loss is the preferred initial fragmentation pathway throughout the alkane series, and that competition from C--C cleavage is strongest for propane.

\begin{figure}[t!]
        \includegraphics[width=1\linewidth]{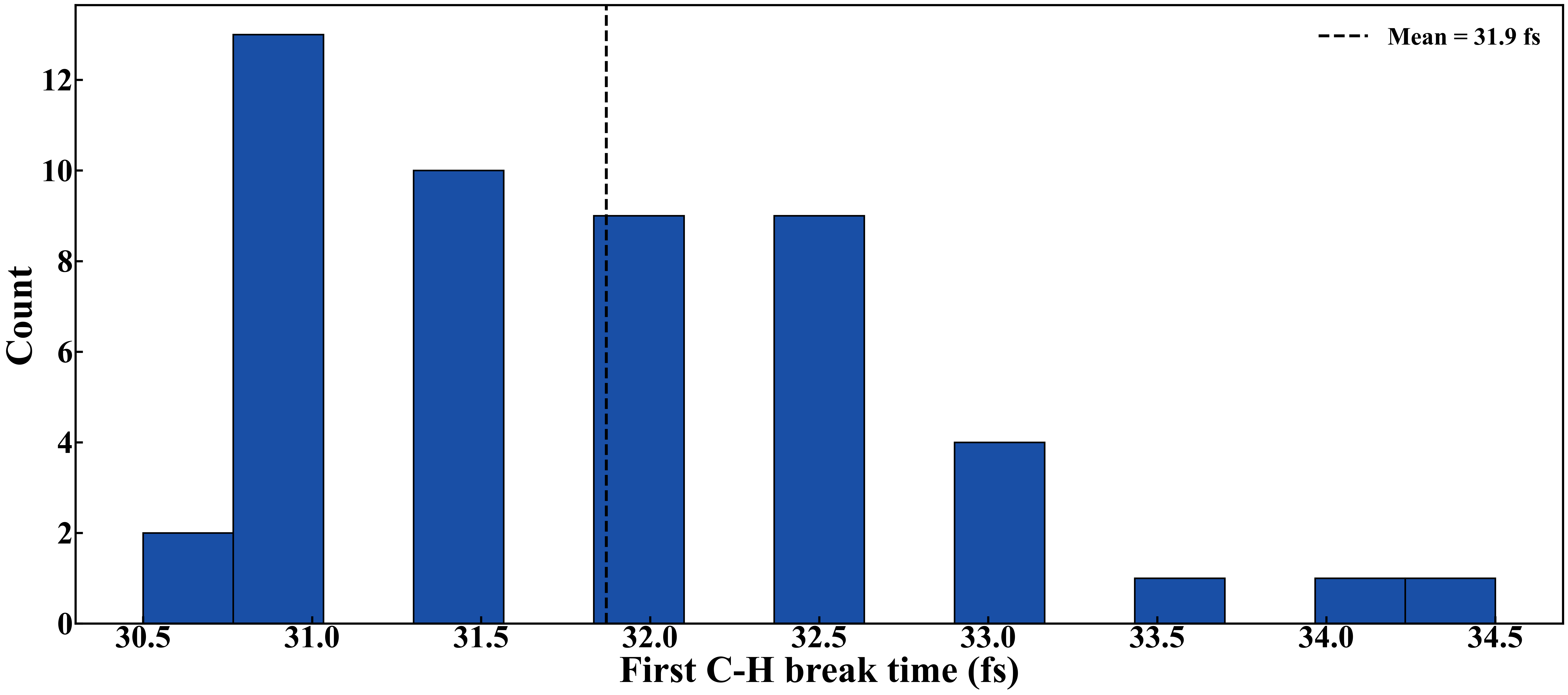}
        \caption{First C--H bond-breaking times. The dashed vertical line marks the mean value, $31.9~\mathrm{fs}$.}
        \label{fig:c2h6_hist_first_CH_break}
\end{figure}

\subsubsection{Ethane $\mathrm{C_2H_6}$}

For $\mathrm{C_2H_6}$, the first C--H bond-breaking times are narrowly clustered near $31$--$32.5~\mathrm{fs}$, with a mean value of $31.9~\mathrm{fs}$, as shown in Fig.~\ref{fig:c2h6_hist_first_CH_break}. Only a small number of trajectories show later C--H rupture, extending to approximately $34.5~\mathrm{fs}$.

\begin{figure*}[!t]
    \centering
    \begin{subfigure}[t]{0.5\textwidth}
        \centering
        \includegraphics[width=\linewidth]{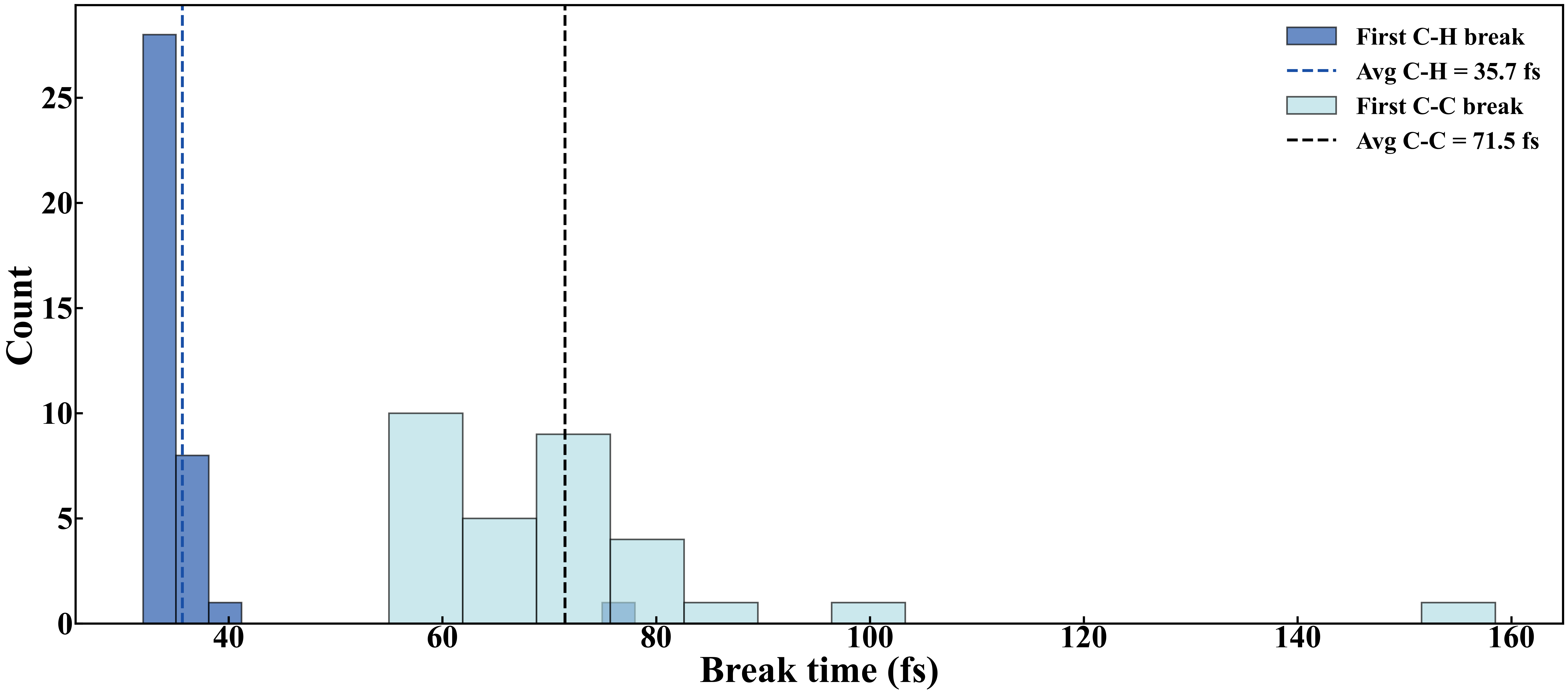}
        \caption{}
        \label{fig:c3h8_his_CH_CC_break}
    \end{subfigure}
    \hfill
    \begin{subfigure}[t]{0.4\textwidth}
        \centering
        \includegraphics[width=\linewidth]{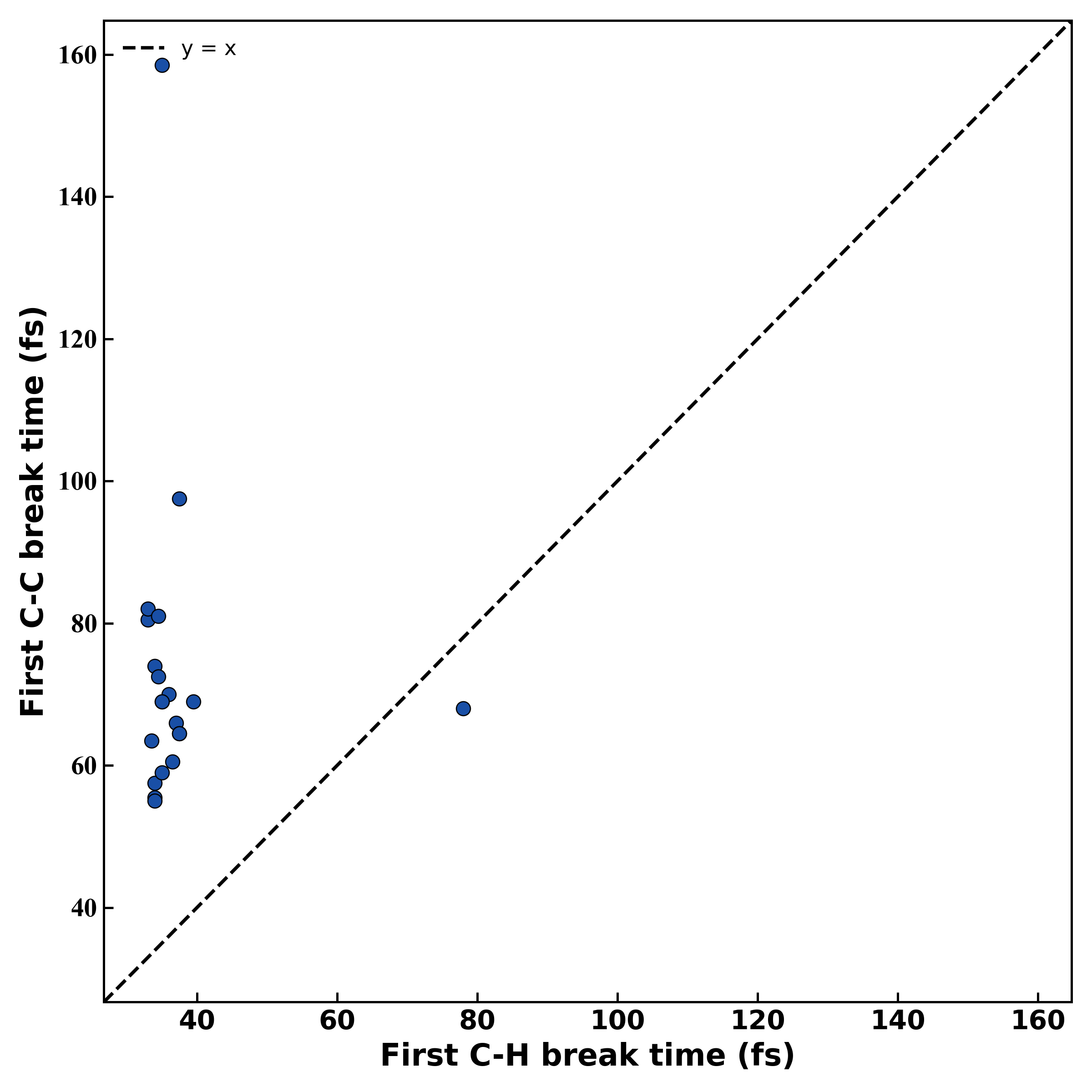}
        \caption{}
        \label{fig:c3h8_scatter_CH_vs_CC}
    \end{subfigure}
    \caption{Bond-breaking dynamics of $\mathrm{C_3H_8}$ under circularly polarized light. (a) Distribution of the first C--H bond-breaking times for $\mathrm{C_3H_8}$. The dashed vertical line marks the mean value, $35.7~\mathrm{fs}$. (b) Correlation between the first C--H and first C--C bond-breaking times for $\mathrm{C_3H_8}$. The dashed line represents $y=x$.}
    \label{fig:c3h8_bond_breaking_dynamics}
\end{figure*}

\subsubsection{Propane $\mathrm{C_3H_8}$}

For $\mathrm{C_3H_8}$, the bond-breaking dynamics are more heterogeneous. The first C--H breaking times are concentrated between approximately $32$ and $38~\mathrm{fs}$, with a mean value of $35.7~\mathrm{fs}$, whereas the first C--C breaking times are more broadly distributed, mostly between $55$ and $85~\mathrm{fs}$, with a mean value of $71.5~\mathrm{fs}$ and one delayed event near $155~\mathrm{fs}$ [Fig.~\ref{fig:c3h8_his_CH_CC_break}]. The correlation plot in Fig.~\ref{fig:c3h8_scatter_CH_vs_CC} shows that most trajectories lie above the $y=x$ line, indicating that C--H rupture usually precedes C--C cleavage. However, several trajectories fall below the diagonal, corresponding to the C--C-first events shown in Fig.~\ref{fig:combined_bar_which_breaks_first}. Propane therefore retains an overall preference for early hydrogen loss, but it exhibits the strongest competition between C--H and C--C cleavage in the present alkane series.

\begin{figure*}[t]
    \centering
    \begin{subfigure}[t]{0.5\textwidth}
        \centering
        \includegraphics[width=\linewidth]{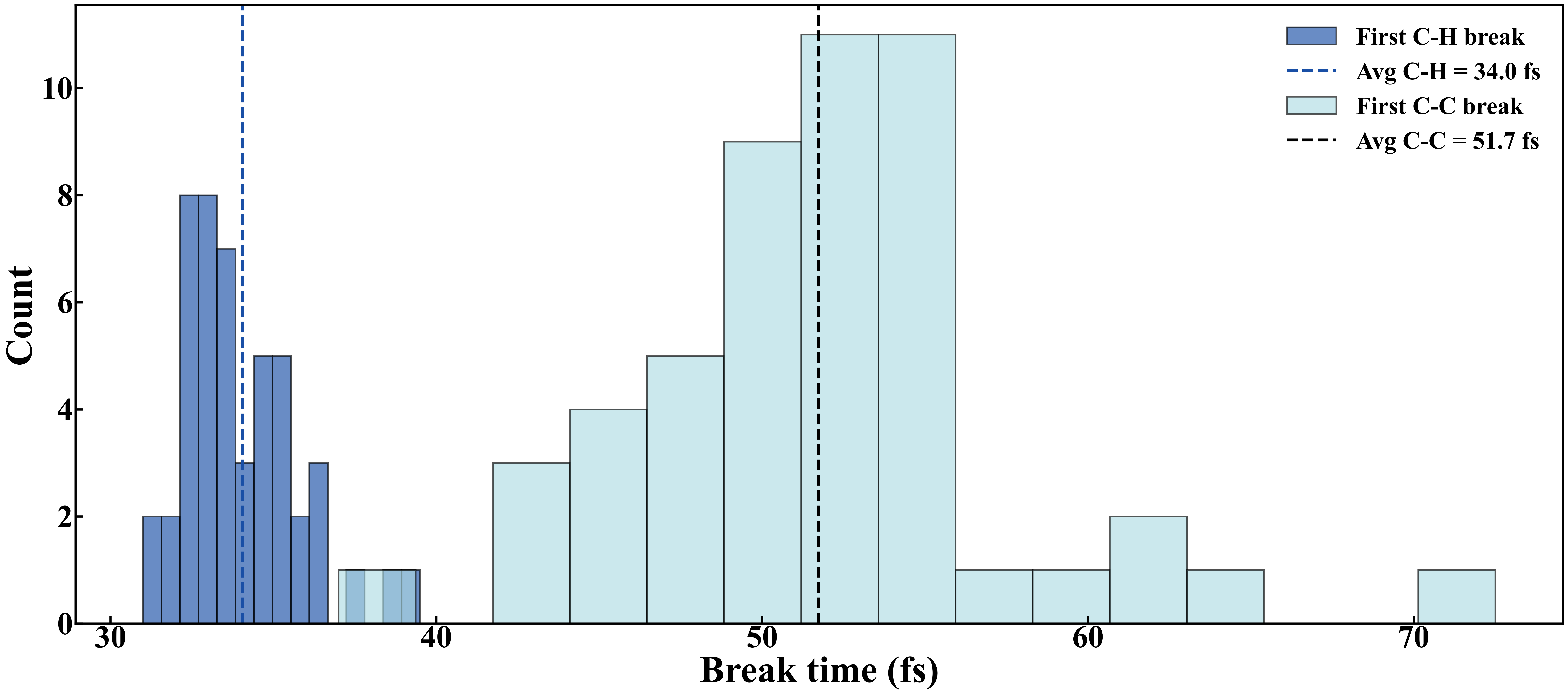}
        \caption{}
        \label{fig:c4h10_his_CH_CC_break}
    \end{subfigure}
    \hfill
    \begin{subfigure}[t]{0.4\textwidth}
        \centering
        \includegraphics[width=\linewidth]{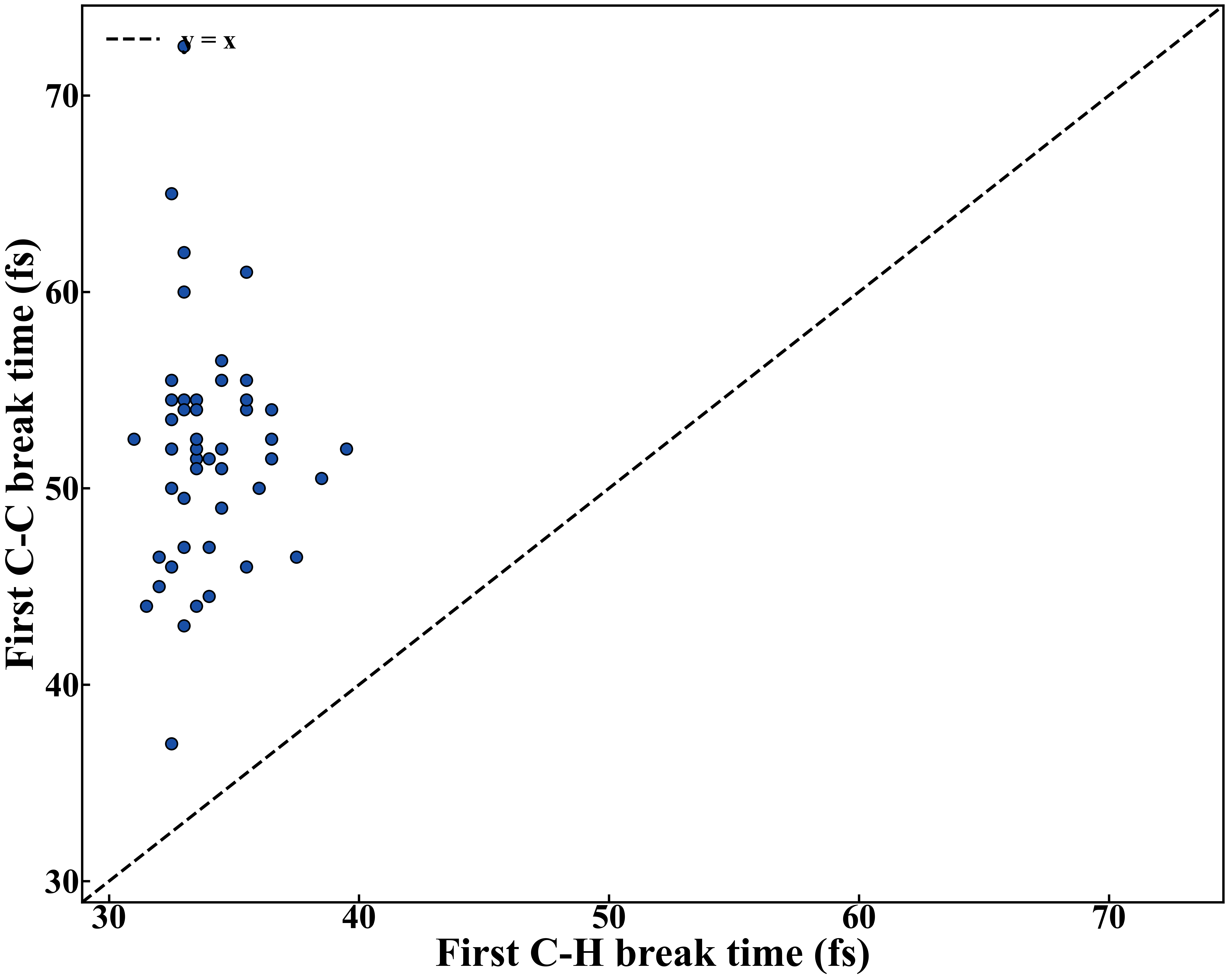}
        \caption{}
        \label{fig:c4h10_scatter_CH_vs_CC}
    \end{subfigure}
    \caption{Bond-breaking dynamics of $\mathrm{C_4H_{10}}$ under circularly polarized light. (a) Distribution of the first C--H bond-breaking times for $\mathrm{C_4H_{10}}$. The dashed vertical line marks the mean value, $34.0~\mathrm{fs}$. (b) Correlation between the first C--H and first C--C bond-breaking times for $\mathrm{C_4H_{10}}$. The dashed line represents $y=x$.}
    \label{fig:c4h10_bond_breaking_dynamics}
\end{figure*}

\subsubsection{\texorpdfstring{$n$-Butane $\mathrm{C_4H_{10}}$}{n-Butane C4H10}}

For $\mathrm{C_4H_{10}}$, the first C--H breaking times are mainly concentrated between $32$ and $36~\mathrm{fs}$, with a mean value of $34.0~\mathrm{fs}$, as shown in Fig.~\ref{fig:c4h10_his_CH_CC_break}. The first C--C breaking events occur later, mostly between $45$ and $56~\mathrm{fs}$, with a mean value of $51.7~\mathrm{fs}$, although a few delayed events extend to about $60$--$65~\mathrm{fs}$, with one event near $72~\mathrm{fs}$ [Fig.~\ref{fig:c4h10_his_CH_CC_break}]. This temporal ordering is also reflected in Fig.~\ref{fig:c4h10_scatter_CH_vs_CC}, where nearly all trajectories lie above the $y=x$ line. Butane therefore remains predominantly C--H-first, but later C--C cleavage contributes to the formation of larger hydrocarbon fragments.

Overall, the bond-breaking analysis reveals a consistent sequence across all three molecules: hydrogen loss occurs first, followed by carbon-backbone fragmentation when it occurs. The main molecule-dependent difference is the degree of competition from C--C cleavage. Ethane exhibits an almost uniform C--H-first pathway with rare delayed backbone cleavage, propane shows the strongest competition between C--H and C--C bond breaking, and butane returns to a predominantly C--H-first pathway followed by later carbon-backbone fragmentation. This trend is consistent with the fragment statistics: ethane mainly preserves the two-carbon backbone, propane displays the broadest set of fragmentation channels, and butane favors larger $\mathrm{C_2}$-based products.

\section{Conclusion}
We have investigated the Coulomb explosion of \(\mathrm{C_2H_6}\), \(\mathrm{C_3H_8}\), and \(\mathrm{C_4H_{10}}\) driven by  circularly polarized laser fields using real-time TDDFT. By comparing circular polarization with linearly polarized fields along the \(x\), \(y\), and \(z\) directions, we find that circular polarization produces the largest ionization for all three molecules under the laser conditions considered here.

The fragmentation and bond-breaking analyses show that hydrogen loss is the dominant nuclear response throughout the alkane series. 
Ethane mainly undergoes dehydrogenation while preserving the two-carbon backbone, propane shows the broadest fragmentation 
landscape and the strongest competition between C--H and C--C cleavage, and butane favors pathways involving relatively 
stable \(\mathrm{C_2}\)-based fragments. The first-bond-breaking analysis further confirms that C--H bond breaking is the 
preferred initial event for all three molecules, while the role and timing of C--C cleavage vary across the series.

Overall, these results show that circular polarization influences
multiple interconnected stages of Coulomb explosion, enhancing early
ionization, lowering the fragmentation threshold, and modifying the
subsequent sequence of bond breaking and fragment formation.
Although the finite trajectory ensemble and the use of ALDA limit the quantitative interpretation of individual 
fragment charges, the main trends are consistent across the three molecules. The present study 
therefore provides a basis for understanding polarization-dependent strong-field fragmentation 
in small hydrocarbons and motivates future work on molecular orientation effects, 
larger hydrocarbon systems, and more detailed comparisons with experimental charge-state-resolved fragment distributions.

\section{Acknowledgment}
This work was supported by the National Science Foundation (NSF) under Grant No. DMR-2217759.
Computational resources were provided by Expanse at the San Diego Supercomputer Center and 
ACES at  Texas A$\&$M University through allocation PHY250153 from the 
Advanced Cyberinfrastructure Coordination Ecosystem: Services $\&$
Support (ACCESS) program, 
supported by NSF grants 2138259, 2138286, 2138307, 2137603, and 2138296. ~\cite{Expanse2025}

\section*{Data Availability Statement}
All data and code are available at https://github.com/kvvandy/tddft.


%

\end{document}